\documentstyle[12pt,math,epsfig]{article}
\renewcommand{\theequation}{\arabic{section}.\arabic{equation}}
\renewcommand{\Re}{{\rm Re}\,}
\renewcommand{\Im}{{\rm Im}\,}
\newcommand{\be}{\begin{equation}}
\newcommand{\ee}{\end{equation}}
\newcommand{\bea}{\begin{eqnarray}}
\newcommand{\eea}{\end{eqnarray}}
\newcommand{\co}{\; \; ,}

\newcommand{\scs}{\co \;}
\newcommand{\sem}{ \; \; ; \;\;}
\newcommand{\per}{ \; .}

\newcommand{\MeV}{\mbox{MeV}}

\newcommand{\Pint}
{\mbox{\scalebox{2}[1]{\rotatebox[origin=c]{50}{--}}\hspace{-2.6ex}$\d\int$}}
\newtheorem{prop}{Proposition}
\newcommand{\spi}{\hspace{-1.3mm}4}
\newcounter{zahler}
\begin{document}

\renewcommand{\thefootnote}{\fnsymbol{footnote}}

\thispagestyle{empty}
\begin{titlepage}
\begin{flushright}
BUTP--99/2\\
\end{flushright}
\vspace{1cm}
\begin{center}
{{\Large{\bf{{One--channel Roy equations revisited\footnote{Work
supported in part by Swiss National Science Foundation} }}}}}

\vspace{1cm}

 J.~Gasser$^{1}$  and
G. Wanders$^2$  \\[1cm]
$^1$ Institut f\"ur  Theoretische Physik, Universit\"at  Bern,\\
Sidlerstrasse 5, CH--3012 Bern, Switzerland \\
gasser@itp.unibe.ch\\[.5cm]
$^2$ Institut de Physique Th\'eorique, Universit\'e  de Lausanne,\\
CH--1015 Lausanne, Switzerland\\
gerard.wanders@ipt.unil.ch\\[1cm]
March 1999 \\[1cm]

{\bf{Pacs:}} 11.30.Rd, 11.55.Fv, 11.80.Et, 13.75.Lb

{\bf{Keywords:}} Roy equations, Dispersion relations, Partial wave analysis,
Meson-meson interactions, Pion-pion scattering, Chiral symmetries

\end{center}

\vskip.5cm

\begin{abstract}
\noindent
The Roy
equation in the single channel case is a nonlinear, singular  integral
equation for the phase shift in the
low--energy  region.
 We first
investigate the  infinitesimal neighborhood of
a given solution, and then
 present explicit
 expressions for amplitudes that satisfy the  nonlinear
equation exactly.
These amplitudes contain free parameters that render  the
non--uniqueness of
 the solution manifest. They
 display, however,  an unphysical  singularity at the upper end
 of the interval considered. This singularity disappears and uniqueness
is achieved if one uses analyticity properties of the amplitudes that
are not encoded in the Roy equation.
\end{abstract}

\end{titlepage}

\newpage
\renewcommand{\thefootnote}{\arabic{footnote}}
\setcounter{footnote}{0}
\setcounter{page}{2}

\section{Introduction\label{in}}

The elastic $\pi\pi$ amplitude has recently been evaluated in the framework
of chiral perturbation theory \cite{chpt} to two loops \cite{2loops}.
The representation is valid in the low--energy region, where the
centre--of--mass energy  of the pions is less than about 400
\MeV.  On the
other hand, precise experimental data are presently  only available
above $\sim$ 600 \MeV.
 In
order to connect the two regions, one may rely on a set of
dispersion
 relations for the partial wave amplitudes due to Roy \cite{roy}.
These allow one to extrapolate the data down to threshold
and to merge with the chiral expansion \cite{coletal}. In the
present article, we investigate this extrapolation procedure in the
 one--channel case.

Roy's representation \cite{roy} for the partial wave amplitudes
 $t^I_l$ of $\pi\pi$ scattering reads as follows,
\bea\label{intro1}
\Re{t}^I_l(s)=k^I_l(s)+ \sum_{I'=0}^2\sum_{l'=0}^\infty
\Pint_{\hspace{-1.3mm}4}^\infty dx\,
 K^{II'}_{ll'}(s,x) \:{\mbox{Im}}\, t^{I'}_{l'}(x)
\sem 4\leq s\leq 60\co
\eea
where $I$ and $l$ denote isospin and angular
momentum, respectively\footnote{We express
all energies in units of $M_\pi$. Further, $\Pint$ denotes a principal
value integral.}.
 The  linear subtraction polynomials $k_l^I$
 are expressed in terms of  the two S--wave
scattering lengths. The kernels
$K^{II'}_{ll'}$
contain
 a diagonal, singular Cauchy kernel that
  generates the right--hand cut in the partial wave amplitudes, as well
 as a logarithmically singular piece that  generates the left--hand cut.

The relations (\ref{intro1})  are consequences of the exact
analyticity properties of the $\pi\pi$ scattering amplitude, of the
Froissart bound and of crossing symmetry. They demonstrate that the
scattering amplitudes are fully determined by the imaginary parts of the
partial waves, except for the two scattering lengths that play the
role of subtraction constants.
Combined with unitarity, (\ref{intro1}) amounts
   to an infinite system of coupled, singular integral equations
 -- the Roy equations -- for
the phase shifts in a low--energy  interval from
threshold $s=4$ to a matching point $s_0 < 60$. In this framework,  the
phase shifts  above the matching point $s_0$,
 the  absorption parameters and the two S--wave
scattering lengths are assumed to be externally assigned.
The mathematical problem consists in solving the Roy equations with
 this input.

Soon after the original article of Roy \cite{roy} had
appeared, extensive
 phenomenological applications were performed \cite{applicroy},
 resulting in a detailed analysis and exploitation of the then available
 experimental data on $\pi\pi$ scattering.
 For a recent review of those results, we refer the reader to
 the article by Morgan and Pennington \cite{penndafne}.
 Parallel to these phenomenological applications, the very
structure of the Roy equations was investigated. In \cite{mahoux},
 extensions of the equations~(\ref{intro1}) were
presented,
valid in the larger range $-28\leq s \leq 125.31$.
 Further,  the manifold of solutions of Roy's equations
 was   investigated as well, both in the single channel
\cite{pomponiuw,n/d,slim}
 as well as in the coupled channel case \cite{epelew}.
 In the late seventies, Pool \cite{proofpool}
  provided a proof that the original, infinite set of integral
  equations does have at least one  solution for $s_0 < 23.31$,
 provided that the driving terms are not too large,
 see also \cite{proofheemskerk}.
Heemskerk and Pool also examined  numerically the solutions of  the
Roy equations, both by solving
 the $N$ equation \cite{proofheemskerk}
 and by using an iterative method \cite{heemskerk}.

It emerged from these investigations that -- for a given input of
absorptive parts, absorption parameters and S--wave scattering
lengths -- there are in general many possible solutions to the Roy equations.
This non--uniqueness is due to the singular Cauchy kernel
 in  the right--hand side of  (\ref{intro1}).
 In order to investigate the uniqueness properties of the Roy system,
 one
 may -- in a first step --  keep only this part of the kernels, as a
result of which the
integral equations decouple: one is
left with a
single channel problem, i.e. a single partial wave, that has, moreover,
 no left--hand cut. This  mathematical problem  was
 examined by Pomponiu and Wanders \cite{pomponiuw}.
 Investigating the
infinitesimal neighborhood of a given solution, they found that the
multiplicity of the solution increases by one whenever the value of
the phase shift at the matching point goes through a multiple of  $\pi/2$.
By
contrast, the number of parameters in the usual partial wave equation
increases in general by two whenever the phase shift at infinity
 passes through a positive integer of $\pi$,
 see e.g. \cite{lovelace,brander}
and references cited therein.

Atkinson and Warnock \cite{n/d} investigated
 a one--channel problem by using $N/D$
methods, that do not require linearization.
Their work may be summarized as follows. First, they find that
 solutions of the Roy equations with $\delta(s_0)\geq -\frac{\pi}{2}$
 are members of solution manifolds depending on $m$ parameters, where
 $m$ denotes the integer part of $2\delta(s_0)/\pi$. Second, these
 solutions may be computed through an integral equation in which the
 $m$ parameters appear explicitly. An exception to these statements
 -- whose proof is  rather complex -- could occur if a certain
 Fredholm operator had unit eigenvalue.
Although the relevant integral equations in \cite{n/d}
exhibit the $m$ parameters, it had not been shown that these  are
 really effective: an arbitrarily chosen set may lead to ghosts.

After 1980,  interest in the Roy equations  waned. In recent years, it has
 however become clear that there are good reasons to revive these techniques.
 First, new $K_{l4}$
 experiments are  planned \cite{kl4dafne} or
 are already under way \cite{kl4brook,pocanic}.
 These will provide new information on the low--energy $\pi\pi$ scattering
 amplitude. A reliable analysis of  available
 low--energy data and high--energy phase shifts  should then again be
 based on Roy equations.
 Second, one can  establish at this point contact with CHPT,  that
will allow one \cite{chpt,stern}
 to gain insight into the structure of the
underlying theory, the Standard Model. First steps in this program
have already been performed \cite{patarakin,buett}.
Of course, this enterprise  requires that one understands  the
structure of the
 Roy equations in all details. It is the aim of this and a following
 article \cite{paper2} to fill existing gaps in this respect and
to provide insights into the problem from an actual point of view, see
 also Ref.~\cite{anant}.

We concentrate  in this work on the uniqueness
properties of the solutions in the one--channel case, and on their
singularity structure.
We start by analyzing the Roy equation with the linearization method
proposed in \cite{pomponiuw} and show  how the emerging
integral equation can be solved by transforming it into a homogeneous
Hilbert problem.
This method is very efficient in determining the multiplicity of
solutions. It does, however,  assume the existence of a
solution. Furthermore, it is not
easy to show that  each solution of the linearized equation
approximates a solution  of the nonlinear one.
For these reasons, we also
 investigate the Roy equation for a
special class of inputs, that allow us to find  explicit solutions. These
solutions do contain parameters that  exhibit their
non--uniqueness.
We then investigate the role of the unphysical singularity that shows up
in the solutions. This singularity manifests itself as a cusp
 in the real and imaginary parts of the amplitude at the matching point.
By making use of analyticity
properties of the amplitude that are not explicit in the Roy
equation,
 we find
that there is a unique solution, devoid of cusps.

Our article is organized as follows. In section 2, we
formulate the mathematical problem we are concerned with here.
In section 3, we analyze the infinitesimal neighborhood of a
solution. The construction of a second exact solution -- once a first
solution is known -- is reduced to a linear problem in section 4.
This allows us to give in section 5  a class of
explicit exact solutions.
In section 6 we show how uniqueness is obtained by use of analyticity
 properties of the amplitudes that are not encoded in the Roy equation.
 A summary and concluding remarks are given in
section 7.
 Details on  precise mathematical formulations are relegated to
appendix A,  where we also construct the general solution of the
linearized problem of section 2.
Appendix B provides the connection with the $N/D$ method \cite{n/d},
and
appendix C contains the proof of the
uniqueness property for an analytic input.

In order to help orient the reader, we note that
the key  statements in this article   are summarized in
{propositions} 1, 2, 3 and 4 in sections 3, 4 and 6.
 Propositions 1  and 3
were established long
 ago in Refs.~\cite{pomponiuw} and \cite{zimmermann}, respectively,
whereas  {propositions} 2 and 4 are -- to the best of our knowledge --
new results.

\setcounter{equation}{0}

%%%%%%%%%%%%%%%%%%%%%%%%%%%%%
%chapter 2
%%%%%%%%%%%%%%%%%%%%%%%%%%%%%

\section{The one--channel Roy equation}

In order to study the
  non--uniqueness properties of the Roy equation, we keep
  only the diagonal, singular Cauchy kernel in (\ref{intro1}).
 The partial wave relations then  decouple, and the left--hand cut in
  the amplitudes disappears.
  We  therefore first explore
 the set of complex amplitudes  $f: [4,\infty)\rightarrow \C$  with
the following properties:

\noindent
i) In an interval $[4,s_0]$ containing the  threshold $s=4$ and a
matching point $s_0$, the real part is given by a dispersion relation
\begin{equation}\label{2one}
\Re f(s)=a+(s-4){1\over \pi}\Pint_{\hspace{-1.3mm}4}^\infty
{dx\over x-4}{\Im f(x)\over x-s}\per
\end{equation}
%\end{document}
ii)
 The imaginary  part $\Im f$ is a given input function $A$ above
 $s_0$,
\begin{equation}\label{2two}
\Im f(s) =A(s),\qquad s\geq s_0\per
\end{equation}
iii) Elastic unitarity holds  below $s_0$,
\begin{equation}\label{2four}
\Im f(s)=\sigma(s)|f(s)|^2 \scs s\in[4,s_0]\sem
\sigma(s)=[1-4/s]^{1/2}\per
\end{equation}
We relegate  a  precise formulation of the regularity
 properties of $f$ to appendix A and simply note
 that, as a minimal requirement,
 the imaginary part $\Im f$ must be continuous in $[4,s_0]$, in particular,
\begin{equation}\label{2three}
\lim_{s\nearrow s_0}\,\Im\,f(s)=A(s_0)\per
\end{equation}
 Equations~(\ref{2one})--(\ref{2three}) constitute
the mathematical problem that we discuss in the following:
determine the amplitudes
$f$ which verify these equations  for  given scattering
 length $a$
and given absorptive part $A$.  This
one--channel problem   allows detailed analytical and numerical
calculations and provides
 useful insight  into the solutions of  the  coupled system considered
e.g. in
Refs.~\cite{coletal,applicroy,epelew}.

Elastic unitarity allows one to reduce the problem to the
determination of a single  real function on the real interval
$[4,s_0]$, because
$f$ is parametrized by its real phase shift
$\delta$ below $s_0$:
\begin{equation}\label{2five}
f(s)={1\over \sigma(s)}{\rm e}^{{\rm i}\delta(s)}\sin\delta(s),\qquad
s\in [4,s_0].
\end{equation}
We choose  the normalization of $\delta$  such that it vanishes at threshold,
$\delta(4)=0$. The boundary condition~(\ref{2three}) becomes
\begin{equation}\label{2six}
\sin^2\delta(s_0)=\sigma(s_0)A(s_0).
\end{equation}
The parametrization (\ref{2five}) allows one to write
Eqs.~(\ref{2one}) - (\ref{2four}) as a nonlinear, singular integral
equation for the phase shift $\delta$,
\bea\label{2twenty}
\frac{1}{2\sigma(s)}\sin{(2\delta(s))}&=&a+\frac{(s-4)}{\pi}
\Pint_{\hspace{-1.3mm}4}^\infty
\frac{dx}{x-4}\frac{\omega(x)}{x-s}\scs s\in [4,s_0]\scs\nonumber\\
\omega(x)&=&\left\{\begin{array}{cl}
\sigma(x)^{-1}\sin^2{\delta(x)} &; \;  4 \leq x \leq s_0\\
A(x) & ; \;x \geq s_0 \co
\end{array}\right.
\eea
with boundary condition (\ref{2six}).
We shall refer to  Eqs.~(\ref{2one})--(\ref{2three})
or Eqs.~(\ref{2six}) and (\ref{2twenty}) as {\it Roy equation with input}
$(a,A)$. We assume in the following that this input is non--vanishing.

Once a solution of the Roy equation is known, the real part of the
amplitude above $s_0$ is obtained from the dispersion relation
(\ref{2one}), and $f$ is then defined on $[4,\infty)$.

The above  formulation of the Roy equation is used in the following
section. There exists  an equivalent approach, based on the
 just mentioned fact that the
dispersion relation (\ref{2one}) can be extended to the half axis
$[4,\infty)$. This   implies that the amplitude $f$ is the boundary
value of an analytic function, holomorphic in $\C \backslash
[4,\infty)$. The Roy
equation then amounts to the construction of this analytic function.
This method is described in detail in section 4 and used in
sections 5 and 6.

\begin{figure}[t]
\begin{center}
\epsfig{file=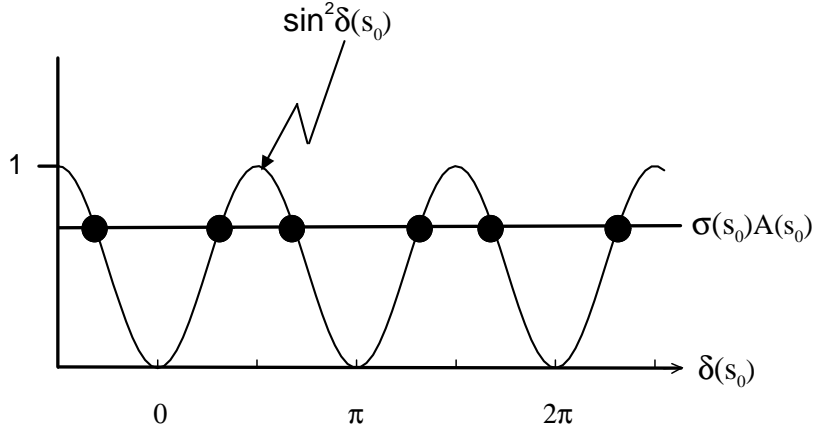,height=10cm,angle=-90}
 \caption{
Solutions to Eq.~\protect{(\ref{2six})}. The filled circles
 correspond to values of the phase shift $\delta(s_0)$ that fulfill
the condition  (\protect{\ref{2six}}).\label{fig1}}
\end{center}
\end{figure}

The value of the phase shift at the matching point
plays a crucial role in the following analysis
\cite{pomponiuw,n/d,epelew}.
 However, the input absorptive part
fixes $\delta(s_0)$ through the boundary condition (\ref{2six})
 only modulo $\pi$ and up to its
sign.  In Fig.~\ref{fig1},
%%%%%%%%%%%%%%%%%%%%%%%%%%%%%%%%%%%%%%%%%%%%%%
%%%%%%%%%%%%%%%%%%%%%%%%%%
\begin{figure}[t]
\begin{center}
\epsfig{file=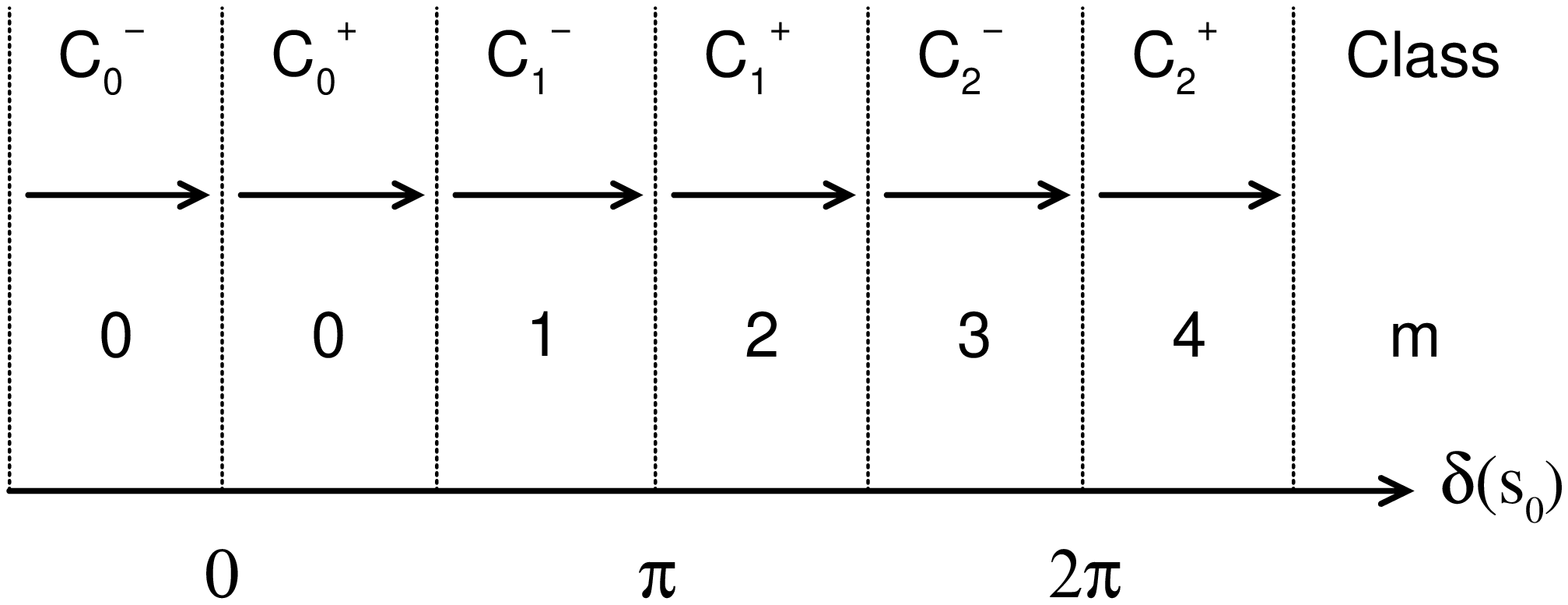,height=10cm,angle=-90}
 \caption{
The classes $C_n^\pm$  as defined in \protect{(\ref{2sevena})}. The
 number $m$ in the lower row  equals the dimension of the
 infinitesimal neighborhood
 of a given  solution, see section 3.
\label{fig2}}
\end{center}
\end{figure}
we display  $\sin^2\delta(s_0)$
as a function of the phase shift $\delta(s_0)$, together with
 the quantity $\sigma(s_0)A(s_0)$, shown with a horizontal
line. The filled circles correspond to  values of $\delta(s_0)$ that fulfill
the condition (\ref{2six}). For reasons outlined in \cite{n/d},
we stick to  phase shifts
 that are not too  negative.
 Furthermore, the case
where the phase
shift at the matching point is a multiple of $\pi/2$ requires special
considerations.  This complication may be avoided with an appropriate
choice of the matching point $s_0$ in actual calculations.
 Unless
stated otherwise, we therefore
 assume in the following that
\bea
\delta(s_0)&>&-\frac{\pi}{2}\co\nonumber\\
\delta(s_0)&\neq& n\pi/2\scs n=0,1,2,\ldots\per
\eea
 The solution of the Roy equation is not unique in general
\cite{pomponiuw,n/d,epelew}, and it is   useful to divide
the manifold of solutions into classes $C_n^{\pm}$,  para\-me\-trized by the
value of the phase shift at the matching point,
\bea\label{2sevena}
C_n^+:&&\delta(s_0)\in (n\pi,n\pi+\frac{\pi}{2})\co \nonumber\\
C_n^-:&&\delta(s_0)\in (n\pi-\frac{\pi}{2},n\pi)\;
 \; ; \; \; n=0,1,2,\ldots \per
\eea
These classes are indicated in figure~\ref{fig2}.
To quote an example, a solution $f$ with $\delta(s_0)=\frac{5\pi}{4}$
belongs to the class $C_1^+$.

Aside from the question of uniqueness, we will be concerned with
 the singularity structure of the solutions: every
 solution of (\ref{2twenty}) with arbitrary input $(a,A)$ is regular on
 $(4,s_0)$,
but singular and only H\"older continuous at the end points of that
 interval.
Whereas the singularity at $s=4$ is due to the threshold behaviour,
 the one at the matching point is unphysical, because the position of
 $s_0$ is arbitrary.
It is only for
 the special class of analytic inputs -- which will be specified in
section 6 --
that there exists a solution that is regular at $s_0$.

Although the original system of coupled, nonlinear and singular
integral equations has been reduced to the simplified equation
(\ref{2twenty}), we are unable to solve it explicitly for an
arbitrary input. For this reason, we have to rely on alternative
tools to achieve our goal.
 In particular,
 we i) investigate the infinitesimal neighborhood of a given solution,
ii) construct explicit solutions in the case where the matching
point is moved to  infinity,
and iii) investigate the manifold of solutions for a conveniently
chosen specific input.
 A clear picture of the multiplicity of the solutions
will emerge in this
 way, and the role of the unphysical singularity
 at the matching point can be investigated in a satisfactory manner.

\setcounter{equation}{0}

%%%%%%%%%%%%%%%%%%%%%%%%%%
%chapter 3
%%%%%%%%%%%%%%%%%%%%%%%%%%

\section{Linearization and existence of ambiguities}

We assume in this section that the Roy equation with input $(a,A)$ does have
a solution $\delta$, and seek for solutions $\delta'$ that are
nearby. These can be determined by linearizing and explicitly solving
the  integral
 equation for the difference of the phase shifts.
As already mentioned in the introduction, this method  has
 two apparent drawbacks: first, one does not  prove the very existence of
 the solution $\delta$. Second, it is not shown that the
$m$--dimensional neighborhood of a solution  is embedded in an
$m$--dimensional manifold.
On the other hand, as we will show in sections 4 and 5, we can
construct explicit pairs $(a,A)$ for which a solution is known and for
which
the ambiguities found below are present. In any case we believe that,
despite its shortcomings, the
linearization method is    very
useful and enlightening, in
particular so in view of its simplicity.

 The solutions of the linearized integral equation show that,  if a
class with $n>0$ is non--empty,  it
contains a continuous family of solutions \cite{pomponiuw}.
 This multiplicity structure  is
identical to the one indicated in  the work of
Atkinson and Warnock \cite{n/d},
 and the investigations of the nonlinear case carried out in later sections
 support this picture.

We now show how this result is obtained when
using the method of \cite{pomponiuw,epelew}. By assumption, both phase
shifts $\delta$ and $\delta'$  satisfy
the integral equation~(\ref{2twenty}).  We wish to determine the difference
\begin{equation}\label{2nine}
\Delta(s)=\delta'(s)-\delta(s).
\end{equation}
Equation~(\ref{2twenty}) for $\delta$ and $\delta'$ results in a nonlinear
singular integral equation for $\Delta$. Assuming $\Delta$ to be small,
linearization of this integral equation  gives
\begin{equation}\label{2ten}
\cos(2\delta(s))h(s)=(s-4){1\over \pi}\Pint_{\spi}^{s_0}dx{1\over x-
4}{\sin(2\delta(x))h(x)\over x-s}\scs s\in[4,s_0]\scs
\end{equation}
where
\begin{equation}\label{2eleven}
h(s)={\Delta(s)\over \sigma(s)}.
\end{equation}
To be consistent with the condition (\ref{2six}),
we require furthermore that
$\delta'(s_0) = \delta(s_0)$, or
\begin{equation}\label{2twelve}
h(s_0)=0\per
\end{equation}
The original Roy equation is   replaced by the
singular linear integral
equation~(\ref{2ten}), to be solved with the boundary
condition (\ref{2twelve}). The latter shows that we can determine in
this manner only those $f'$ which belong to the same class as $f$.

Constructing the solution of (\ref{2ten}) is equivalent to  solving a boundary
value problem for analytic functions -- a so called
Hilbert problem
 \cite{muskhel}. To make this article
self--contained, we discuss  the procedure
in  appendix A, where it is shown  that the general solution of
(\ref{2ten}) is given by \cite{pomponiuw}
\begin{equation}\label{2thirt}
h(s)=(s-4)G(s)P(s),\qquad s\in[4,s_0],
\end{equation}
with
\begin{equation}\label{2fourt}
G(s)={1\over (s_0-s)^m}\exp\left[{2\over \pi}
\Pint_{\spi}^{s_0}dx{\delta(x)\over x-
s}\right]\scs
\end{equation}
and where $P(s)$ is an arbitrary real polynomial of degree $m-1$,
with
\begin{equation}\label{2fift}
m=\left\{\begin{array}{cl}
\left[2\delta(s_0)\over \pi\right] \quad&\mbox{if }
\delta(s_0)>{\pi\over
2},\\[2mm]
0 & \mbox{if }-{\pi\over 2}<\delta(s_0) < {\pi\over 2}.
\end{array}\right.
\end{equation}
The symbol $\left[x\right]$ in (\ref{2fift}) denotes the greatest integer
not exceeding $x$.
Polynomials of negative degree are considered to be identically zero
here and in the following. For $\delta(s_0)>0$ ,
the number $m$ coincides with the so--called
{\it index} of the Hilbert problem. For a monotonically increasing
phase, it counts the number of times
$\delta(s)$ goes through multiples of $\pi/2$ as $s$ varies
 from threshold to the matching point $s_0$. We indicate
in figure~\ref{fig2} some of its values.

The following proposition summarizes the results obtained in this
section and in appendix A.
\begin{prop}\label{prop1}
Let $\delta$ be a solution of
 Eq.~(\ref{2twenty}). It is an isolated solution of that equation if
$-\frac{\pi}{2} < \delta(s_0)<\frac{\pi}{2}$. If $\delta(s_0)>
\frac{\pi}{2}$, the infinitesimal neighborhood of $\delta$ is an
$m$--parameter family of solutions $\delta'$ with
$\delta'(s_0)=\delta(s_0)$, where $m$ is given in (\ref{2fift}).
\end{prop}

%\end{document}
\subsection*{Comments}
\begin{enumerate}

\item
An obvious question concerns the interpretation of the $m$ parameters
 in the polynomial $P$ in (\ref{2thirt}).
 In the case where the phase shift
$\delta$ is monotonically increasing and where $m$ is an even integer,
$f$ exhibits $m/2$ resonances on $(4,s_0)$. The nearby $f'$ has also
 $m/2$ resonances, with slightly modified positions and widths. These
 changes are fixed by the $m$ coefficients in  $P$. This
 is verified in an example that we study in subsection 5.1. The situation
 is not so simple if $m$ is odd. The case $m=1$ is illustrated in
 subsection 5.3.

\item
Atkinson and Warnock \cite{n/d} have studied these
parameters in a   single
channel  Roy equation that has  a  structure which is  similar  to the one
studied here.
 Using an $N/D$--method, they find that, when $m$ is an
even integer, the parameters are related to the position and residues of
the CDD poles between threshold and the matching point.
In the case
where $m$ is an odd integer, one of these parameters is connected with
the  singular nature of the $N$ equation.
\item
The difference $\triangle$ is singular at the matching point. This is
a signal of the singularity of the general solution of the Roy equation
mentioned in section 2.

\item Our discussion deals exclusively with restrictions to the low--energy
interval $[4,s_0]$. Once $\Delta$ is determined on that interval, the value of
$f'$ for all real $s$ is obtained from the dispersion relation
(\ref{2one}).
 In the linear regime,
\begin{equation}\label{eq16}
f'(s)=f(s)+\frac{(s-4)}{\pi}\lim_{\epsilon\searrow 0}\int_4^{s_0}dx
{\sin(2\delta(x))h(x)\over x-4}{1\over x-s-{\rm i}\epsilon}\per
\end{equation}
Consistency with unitarity is then
by no means ensured above $s_0$: if $\sigma f$ stays on or inside the Argand
circle, $\sigma f'$ may well be outside. This shows that physical requirements
which are not encoded in the Roy equation can reduce the
ambiguities.

\item Our method is also applicable if the first inelastic threshold is below
$s_0$, provided that the absorption parameter $\eta$ is known. The
representation~(\ref{2five}) is replaced by
\begin{equation}\label{2sixt}
f(s)={1\over 2{\rm i}\sigma(s)}\left[\eta(s){\rm e}^{2{\rm i}\delta(s)}-
1\right].
\end{equation}
The form of the linearized equation~(\ref{2ten}) is unchanged, whereas $\eta$
enters into the definition of the unknown, $
h(s)={\eta(s)\sigma(s)^{-1}}\Delta(s).$
With this modification, (\ref{2thirt}) remains valid.
\end{enumerate}

\setcounter{equation}{0}

%%%%%%%%%%%%%%%%%%%%%%%%%%
%chapter 4
%%%%%%%%%%%%%%%%%%%%%%%%%%%%%%%%%%%%

\section{The full amplitudes}
\subsection{Matching at infinity}
\begin{figure}[h]
\begin{center}
\epsfig{file=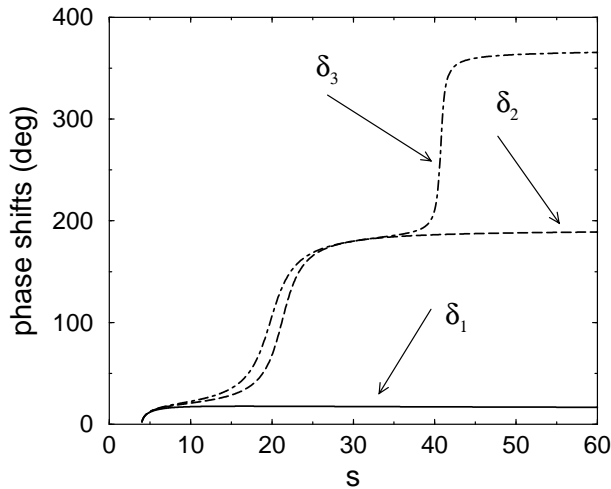,height=8cm,angle=-90}
 \caption{
The phase shifts $\delta_{1,2,3}$ that correspond to the
solutions $f_{1,2,3}$ of
 equation (\protect{\ref{eq4_a1}}), with parameters $a=0.5$,
 $(s_2,r_2)=(30,40),(s_3,r_3)=(45,40).$
\label{fig_1}}
\end{center}
\end{figure}

It is instructive to consider the case where the matching point $s_0$
is moved to infinity, because  the Roy equation can then be solved
 explicitly.
The amplitude vanishes at infinity \cite{brander} in this case,
 as a result of which the input reduces to the scattering length $a$.
 Equation
(\ref{2twenty}) becomes\footnote{A subtraction is in fact not
needed here \cite{brander}. We  stick to the present formulation for an easier
comparison with the Roy equation at a finite matching point.}
\bea\label{eq4_a1}
\frac{1}{2\sigma(s)}\sin{(2\delta(s))}&=&a+\frac{(s-4)}{\pi}\Pint_{\spi}^\infty
\frac{dx}{x-4}\frac{\sin^2{\delta(x)}}
{\sigma(x)(x-s)}\scs s\in[4,\infty]\scs\nonumber\\
\eea
 to be solved for given scattering length $a$. For $a>0$, a solution
is provided by the phase shift of the amplitude
\begin{equation}\label{eq4_a2}
f_1(s)=\left[{1\over a}
-\rho(s)\right]^{-1}\co
\end{equation}
where $\rho$ is the Chew--Mandelstam function
\begin{equation}\label{eq4_a3}
\rho(s)={1\over \pi}\sigma(s)\left\{\ln{1-\sigma(s)\over 1+\sigma(s)}
+i\pi\right\} \sem s\geq 4\per
\end{equation}
In figure~\ref{fig_1}, we display the phase shift $\delta_1$
 of $f_1$ with a solid line  for $a=0.5$.
%%%%%%%%%%%%%%%%%%%%%%%%%%
 This is not the only solution -- further
examples are  e.g. the phase shifts of
\bea\label{eq4_a4}
\frac{1}{f_2}&=&\frac{1}{f_1}+\frac{s-4}{s_2-4}\frac{r_2}{s-s_2}
\scs\nonumber\\
\frac{1}{f_3}&=&\frac{1}{f_2}+ \frac{s-4}{s_3-4}\frac{r_3}{s-s_3}\sem
\,\, s_i>4\scs r_i>0\per
\eea
 The corresponding phase shifts are again
displayed in figure~\ref{fig_1}, for
 $(s_2,r_2)=(30,40); (s_3,r_3)=(45,40).$ We note that
the phase shifts tend to  multiples of $\pi$ at infinity,
 $\delta_i(\infty)=(i-1)\pi,i=1,2,3$.
 [The complications with $\delta(s_0)=n\pi/2$, that we mentioned in
 section 2, disappear when
$s_0=\infty$.]

 These examples show  that the Roy equation with
matching point at infinity allows for many solutions. The poles at
$s_i$ are CDD poles \cite{cdd}, and one has to specify the phase shift
at infinity as well as the CDD parameters $r_i,s_i$ in order to pin
down a solution completely. The unsubtracted version of equation
(\ref{eq4_a1}) is discussed in
detail in \cite{brander}.

\subsection{Matching at finite energy}

We now present  an approach that will allow us
 to construct exact
solutions of the Roy equation with a finite matching point $s_0$.
Although we cannot solve the Roy equation with an arbitrary input, we
 can construct explicit amplitudes $f$ that satisfy the dispersion
 relation (\ref{2one}) and verify elastic unitarity below some
 $s_0$. An amplitude with this property defines an input
 $a_f\doteq f(4)$, $A_f(s)\doteq \Im f(s),s\geq s_0$,
and is itself a solution of the Roy equation with this input. This
holds true e.g. for the amplitudes $f_i$ in
 Eqs.~(\ref{eq4_a2}) and (\ref{eq4_a4}).
 Our goal is to show that -- given $f$ -- one can construct
 other solutions $f'\neq f$, with the same input.
In this manner, we find  that
the  Roy equation  has in general solutions in
different classes $C_n^\mu$. This cannot be seen in the
linearization framework discussed in the last section, since there, one has to
assume that the phase shifts of the original and of the new solution
coincide at the matching point, as a result of which the old
and new solutions stay in the same class.

Let us  describe the procedure in detail. Until now we worked
 with amplitudes that are complex functions of the real
 variable $s \geq 4$. As announced in section 2, we now define our
 amplitudes as  analytic functions in the complex $s$--plane,
 cut along the real axis for $s\geq 4$.
 The corresponding  amplitudes verifying
 Eqs.~(\ref{2one}) -- (\ref{2three}) are the boundary values
 $f_+$ of $f$, defined as
\bea
f_+(s)=\lim_{\epsilon\searrow 0}\,f(s+{\rm i}\epsilon)\,\, ,
\qquad s\in [4,\infty) \per
\eea
In particular, we  consider the set of functions $f$  with the following
properties:
\begin{enumerate}
\item[i)] $f$ is holomorphic in $\C\backslash[4,\infty)$ and verifies the
dispersion relation (\ref{2one}), written for $f_+$.
\item[ii)] $f_+$ is elastic below the matching point,
\bea
f_+(s)=\frac{1}{\sigma(s)}e^{i\delta(s)}\sin{\delta(s)}\co\qquad
 s\in [4,s_0]\per
\eea
\item[iii)] $f_+$ satisfies the regularity requirements listed in
 subsection A.2.
\end{enumerate}
Let $f$ satisfy i)--iii). Its boundary value $f_+$  is a solution of
 the Roy equation
with input
\bea\label{4input}
a_f&\doteq & f(4)\scs\nonumber\\
A_f(s)&\doteq& \Im{f_+}(s)\scs s\geq s_0\per
\eea
 In this and the
following section,
we show how to construct functions $f'\neq f$ that satisfy
i)--iii) with
\bea\label{3two-}\nonumber\eea

\vspace{-2cm}

\addtocounter{zahler}{1}
\renewcommand{\theequation}{\arabic{section}.\arabic{equation}\alph{zahler}}
\bea
 \hspace{-1.5cm}f'(4)=a_f\co\label{3two}
\eea
\addtocounter{zahler}{1}
\addtocounter{equation}{-1}

\vspace{-2cm}

\bea
\Im{f_+'(s)}=A_f(s)\scs s\geq s_0\label{3fourbis}\per
\eea
\setcounter{zahler}{0}
It is clear that $f'_+$ is then also a solution of the Roy equation with
input (\ref{4input}), and the existence of an $f'$ with the above mentioned
properties therefore establishes the
non--uniqueness of the solution of the Roy equation.

In order to  construct an $f'$, we use the following Ansatz,
\begin{equation}\label{3four}
{1\over f'(s)}={1\over f(s)}+(s-4){H(s)\over D(s)}\per
\end{equation}
 Here, $H$ is  an Omn\`es--type function \cite{pomponiuw},
an analogue of $\bar{G}$ defined in
 (\ref{a3b}),
\begin{equation}\label{3five}
H(s)=\left({s_0\over s-s_0}\right)^m\exp\left[-{2\over
\pi}s\int_{s_0}^\infty{{\rm d}x\over x}{\theta(x)\over x-s}\right]\co
\end{equation}
with $m$ given in (\ref{2fift}).
The function $\theta$  is H\"{o}lder continuous and
equal to ${\rm arg}\,f_+$ modulo $\pi$, with $\theta(s_0)=\delta(s_0)$.
(We assume here in addition that  $\Im\,f_+\geq 0$. As a result of
this, we can define
${\rm arg}\,f_+$ such that $0\leq {\rm arg}\,f_+\leq\pi$.)
 The function $D$ is  meromorphic\footnote{In order to simplify
the presentation, we shall -- without any further mention --  use the fact
that  all our analytic functions satisfy
 $F(s)=\overline{F(\overline{s})}$.}
 in
$\C\backslash[s_0,\infty)$ -- which ensures that ii) is fulfilled -- and
has to be constructed such that also the remaining
conditions are satisfied.
We write
\bea\label{3seven}
D=D_1+D_2\co
\eea
where $D_1$ is a  meromorphic component of $D$, and where $D_2$ is
  regular in  $\C\backslash[s_0,\infty)$. Condition (\ref{3two})
  requires $D(4)\neq 0$.
 The definition of $H$ has been  chosen
such that the condition
(\ref{3fourbis}) amounts to a simple linear constraint
on $D_2$,
\bea\label{3eight}
\Im \, D_{2+}(s)&=&\mu (s)\scs s\geq s_0\co\nonumber\\
\mu (s)&=&(s-4)|H(s)|A_f(s)\per
\eea
Once this condition is fulfilled and $D(s_0)$ is not zero, $\Im \, f'$
is continuous at $s_0$ if $H(s_0) = 0.$ This holds true if
$\delta(s_0)>0$. Choosing a particular function $D_2$ verifying the
condition (\ref{3eight}), the arbitrariness in $f'$ is entirely
contained in $D_1$. This function has to be such that $f'$
satisfies i) and iii). Therefore, $1/f'$ and $1/f_+'$ have to be nonzero
 on $\C\backslash[4,\infty)$ and  on $[4,\infty)$, respectively.
 The non--uniqueness of the solution of the Roy equation is due to the very
 existence of such functions $D_1$ --
 we  provide explicit examples in the following
section.
Our discussion leads to \sloppy
\begin{prop}\label{prop2}
 Let $f$  be an amplitude verifying
conditions i)--iii), with $\Im{f_+}\geq 0,$
and $\delta(s_0)> 0$. Let $f'\neq f$ also  verify i)--iii),
together with
the  conditions (\ref{3two-}). Then $f'$  can be written in the
form (\ref{3four}) -- (\ref{3seven}),
where $D_1$ is  a meromorphic component of $D$, and  $D_2$ is
regular in $\C\backslash[s_0,\infty)$, with spectral function (\ref{3eight}).
\end{prop}\fussy
In the $N/D$ approach \cite{n/d}, an important part of the
non--uniqueness is due to  arbitrariness in the CDD poles. Their
relation with the function $D$ in Eq.~(\ref{3seven})  is explained in
appendix B.

\setcounter{equation}{0}

%%%%%%%%%%%%%%%%%%%%%%%%%%
%chapter 5
%%%%%%%%%%%%%%%%%%%%%%%%%%%

\section{Explicit exact solutions}

\begin{figure}[h]
\begin{center}
\epsfig{file=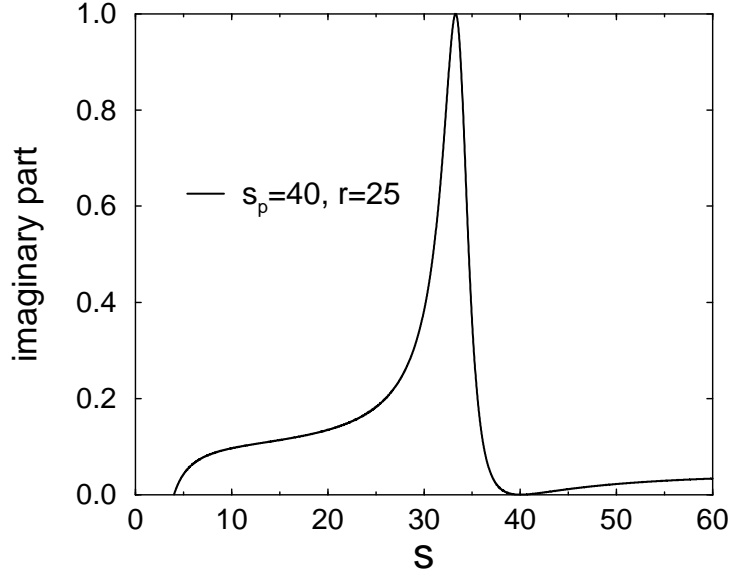,width=8cm,angle=-90}
 \caption{ The imaginary part of the function $\sigma f$
 in (\protect{\ref{3ten}}).
The parameters used
 are the ones  in (\protect{\ref{eq51}}). The phase shift $\delta$ of $f$ is
 displayed with a solid line in Fig.~\protect{\ref{fig4}}.
 \label{fig3}}
\end{center}
\end{figure}

Here, we illustrate the procedure described in subsection 4.2
with specific examples.
An amplitude fulfilling the conditions i)--iii) of that subsection  can be
written as follows,
\bea\label{eq5one}
f(s)=\left[\frac{1}{a}+(s-4)\phi(s) -\rho(s)\right]^{-1}\, ,
\eea
where $\phi$ is a suitable  function  that is meromorphic
in the complex  $s$--plane, cut along the real axis for  $s\geq s_0$.
 The amplitudes $f_i$ in Eqs.~(\ref{eq4_a2}) and
 (\ref{eq4_a4}) have this form with rational $\phi$. To
 keep our calculations simple, we work with the resonant amplitude
 $f_2$ in (\ref{eq4_a4}).
 We drop  indices and write
\begin{equation}\label{3ten}
f(s)=\left[{1\over a}+{s-4\over s_p-4}{r\over
s-s_p}-\rho(s)\right]^{-1}\co
\end{equation}
where $s_p>4$, $r>0$, $a>0$.
 As $f$ is elastic on $[4,\infty)$, the argument $\theta$ in
(\ref{3five}) coincides with the phase shift $\delta$.
 In figures~\ref{fig3} and \ref{fig4}, we
display  the behaviour of   $f$ in the case where
\bea\label{eq51}
a=0.5\co\;\; s_p=40,\;\; r=25\per
\eea
The imaginary part of $\sigma f$ is shown
in Fig.~\ref{fig3}, whereas its phase shift $\delta$  is indicated
 with a solid line in Fig.~\ref{fig4}.
Note that the matching point
$s_0$ does not occur in $f$ -- it may be chosen at one's
convenience.
 We observe that $s_p$ becomes  a CDD
 pole -- as defined in \cite{n/d} -- if  $s_p< s_0$.

In the following, we keep the amplitude $f$ fixed and
bring it into class $C_1^+$ or $C_1^-$ by an appropriate choice of $s_0$.
 To illustrate, for $s_0=50$, $f$ is in class $C_1^+$, and $s_p$ is a
CDD pole.

We notice that $f$ is a solution of the Roy equation with input
$(a_f,A_f)$ which is regular at $s_0$. We are thus dealing with one of
the special inputs mentioned in section 2.

%%%%%%%%%%%%%%%%%%%%%%%%%%

\subsection{Shift and suppression of a resonance: $f\in C_1^+$,
$f'\in C_1^+,C_0^+$}

We place $s_0$ above $s_p$,
 as a result of which $\pi<\delta(s_0)<3\pi/2$, such that $f$ belongs to
 class $C_1^+$.
We first construct amplitudes $f'$ in the same class $C_1^+$.
This implies that $\delta'(s_0)=\delta(s_0)$. We shall find that,
qualitatively, $f'$ is obtained from $f$ by a shift of the position and a
change of the width of its resonance. As $\delta'(s_0)\in(\pi,3\pi/2)$, $1/f'$
has a single pole $s'_p$ on $(4,s_0)$. The second term in the Ansatz
(\ref{3four}) has to cancel the pole
of $1/f$ at $s_p$ and replace it by a new pole at $s'_p$.

It is convenient to redefine $D_1$ and $D_2$ in (\ref{3seven}) by
writing
\begin{equation}\label{3twelve}
D(s)=(s-s_p)(s-s'_p)[D_1(s)+D_2(s)]\co
\end{equation}
where $D_1$ is meromorphic, and
\begin{equation}\label{3thirt}
D_2(s)={1\over \pi}\int_{s_0}^\infty{\rm d}x{\mu(x)\over (x-s_p)(x-
s'_p)}{1\over x-s}\per
\end{equation}
The integral converges because $\mu(x)=O(x)$ at infinity. We require
regularity of $1/f'$ at $s_p$ and fix the residue $r'$ of its pole at
$s'_p$, with
\bea\label{5conditions}
s_p'\neq s_p\scs 4 < s_p,s_p' < s_0\scs r' >  0\per
\eea
 This gives two conditions which completely determine a two--parameter
Ansatz for $D_1$. One finds that the adequate Ansatz is
\begin{equation}\label{3fourt}
D_1(s)={1\over \alpha s+\beta}\per
\end{equation}
The two constraints on $D_1$ give
\begin{equation}\label{3fift}
\alpha={\bar{r}'\over R'}-{\bar{r}\over R}\co \qquad
\beta={\bar{r}s'_p\over R}-
{\bar{r}'s_p\over R'}\co
\end{equation}
with
\begin{equation}\label{3sixt}
\begin{array}{ll}
\displaystyle \bar{r}={r\over (s_p-4)H(s_p)},& \displaystyle \bar{r}'={r'\over
(s'_p-4)H(s'_p)},\\[6mm]
R=1+\bar{r}(s_p-s'_p)D_2(s_p),&R'=1+\bar{r}'(s_p-s'_p)D_2(s'_p)\per
\end{array}
\end{equation}
The function $H$ is obtained from (\ref{3five}) with $m=2$.
Notice that $\alpha$ and $\beta$ are small if the pair $(s'_p,r')$ is close to
$(s_p,r)$. In particular, let
\begin{equation}\label{3sevent}
\epsilon = {\mbox {max}}\left({|s'_p-s_p|\over s_p}\scs{|r'-r|\over r}
\right)\co
\end{equation}
with $\epsilon$ small.  We then have
\begin{equation}\label{3eightt}
\alpha=O(\epsilon),\qquad \beta =O(\epsilon)\per
\end{equation}
Inserting the expressions (\ref{3twelve}) and (\ref{3fourt}) into equation
(\ref{3four}) we get
\begin{equation}\label{3ninet}
{1\over f'(s)}={1\over f(s)}+{(s-4)H(s)\over (s-s_p)(s-s'_p)}{\alpha
s+\beta\over 1+(\alpha s+\beta)D_2(s)}\per
\end{equation}
We see that $f'$ is regular in $\C\backslash[4,\infty)$ and is bounded on
its cut if $1/f'$ has no zero. Eqs.~(\ref{3ninet}) and (\ref{3eightt}) show
that $1/f'$ is close to $1/f$ if $\epsilon$ is small and $s$ is outside the
vicinity of $s_p$ and $s'_p$ and not too large.
 The inverse amplitude $1/f'$
is nonzero for such values of $s$ because $1/f$ is nonzero. One finds that
$1/f'$ is also nonzero near $s_p$ and $s'_p$. Since $D_2$ and $H$
behave at infinity as
$s^{-1}\ln|s|$ and $\ln^2|s|$, respectively,  the second term
in (\ref{3ninet}) is no longer a
small correction when $|s|$ becomes large, and a detailed analysis is needed.
It is relatively easy to see that no unwanted zeros show up if $\epsilon$ is
small and $\beta/\alpha$ is not too large. Consequently we are sure that
under these
conditions (\ref{3ninet}) provides a two--parameter family of solutions of
the Roy equation.
 This example provides an illustration of the $N/D$ framework outlined in
appendix B. In that language,  $s_p$ and $s_p'$ correspond to zeros $z_j$.
 The CDD pole of $f$ at $s_p$ is removed and replaced by a new one at $s_p'$.

\begin{figure}[h]
\begin{center}
\epsfig{file=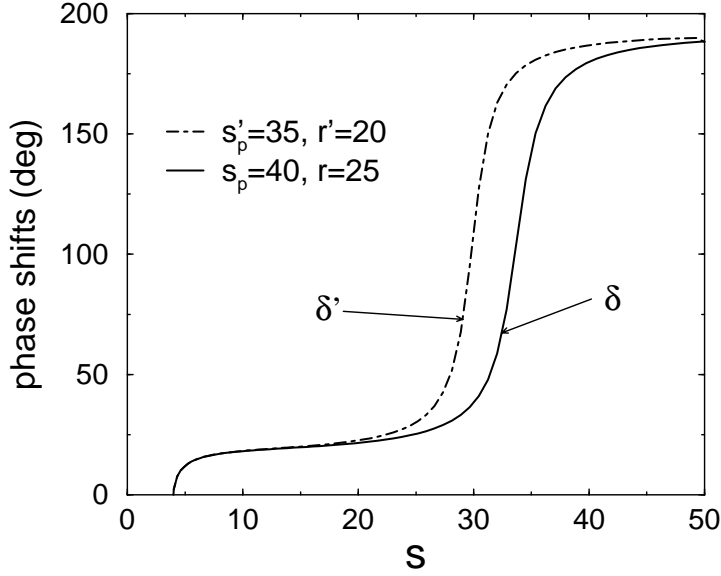,width=8cm,angle=-90}
 \caption{A shift in the class  $C_1^+$. The solid
 line corresponds to the phase shift $\delta$ of the function  $f$ shown in
 Fig.~\protect{\ref{fig3}}, whereas
 the dot--dashed line displays the phase $\delta'$ of $f'$, evaluated from
 (\protect{\ref{3ninet}}),
with $\alpha,\beta$ calculated from (\protect{\ref{3fift}$,$
 \ref{3sixt}}), using the parameters \protect{(\ref{5new2})}.
\label{fig4}}
\end{center}
\end{figure}

We have numerically verified that $\epsilon$ need not be small for the
above conclusions to hold.
 In particular, there is a finite interval for $\epsilon$ such that
 i) there are no additional
singularities in $f'$, and ii) the
dispersion relation (\ref{2one}) is fulfilled. To illustrate, we
choose
\bea\label{5new2}
s_p'=35\scs r'=20\scs s_0=50\per
\eea
In this case one has, using for $f$ the parameters displayed in
(\ref{eq51}),
\bea
\frac{|s_p'-s_p|}{s_p}=0.125\scs \frac{|r'-r|}{r}=0.2\per
\eea
These quantities are not small.
Nevertheless, $f'$ is a
solution of the Roy equation in class $C_1^+$.
We display
 the phase shifts of $f$ and $f'$ in Fig.~\ref{fig4} with
a solid and dot--dashed line, respectively.
As required, the two phase shifts
 agree at the matching point $s_0$ --
 the parameters (\ref{5new2}) really  correspond to a shift
 $C_1^+\rightarrow C_1^+$.

%%%%%%%%%%%%%%%%%%%%%%%%%%
An expression for $f'$ can be obtained in the linear regime from the solution
(\ref{2thirt}) of the problem on $[4,s_0]$. It coincides with the result
(\ref{3ninet}) to first order in $\epsilon$, if the polynomial
in (\ref{2thirt}) is
\begin{equation}\label{3twen}
P(s)=C[(\bar{r}'-\bar{r})s+\bar{r}s'_p-\bar{r}'s_p]\co
\end{equation}
where $C$ is a constant determined by the phase shift $\delta$. This
establishes the relation with the coefficients in the arbitrary polynomial $P$
in our example and provides an interpretation of these coefficients.

The method allows the construction of solutions $f'$ which are no longer
resonant. This is simply achieved by setting the residue $r'$ to zero.
Equation~(\ref{3fift}) then gives $\alpha s+\beta=-\bar{r}(s-s'_p)/R$, and
$1/f'$ becomes
\begin{equation}\label{3twone}
{1\over f'(s)}={1\over a}+\bar{r}{s-4\over s-s_p}\left[H(s_p)-{H(s)\over
L(s)}\right]-\rho(s)\co
\end{equation}
where
\begin{equation}\label{3twtwo}
L(s)=1-\bar{r}\left[(s-s_p)D_2(s)+(s_p-s'_p)\left(D_2(s)-
D_2(s_p)\right)\right]\per
\end{equation}
As $1/f'$ has no pole on $[4,s_0]$, $\delta'(s_0)=\delta(s_0)-\pi$,
and $f'\in C_0^+$. We have to
make sure that $1/f'$ has no unwanted zeros. The conclusions reached in the
case $r'\neq 0$ apply here and $1/f'$ is non--vanishing if the residue $r$ is
small, i.e.~if the resonance in $f$ is narrow. On also finds that $\Re(1/f')$
does not have a zero on $[4,s_0]$ either. Therefore the phase shift $\delta'$
stays below $\pi/2$, and $f'$ is non--resonant. It is easy
to see that $L$ does not depend on the choice of $s'_p$. This means that the
amplitude $f'$ defined in (\ref{3twone}) effectively contains no free
parameter. This amplitude is the unique solution of the Roy equation with
input (\ref{3two-}) belonging to class $C_0^+$. This uniqueness is in
accordance with the results of section~3. In the $N/D$    language,
the uniqueness stems from the fact that $f'$ has no CDD pole, and
$\delta'(s_0) < \pi/2$.
What is new with respect to section~3 is that our example shows
 explicitly that the same Roy
equation has solutions in different classes $C_n^\pm$. At given input
$(a,A)$, the number of resonances below $s_0$ is not fixed unless one
imposes the precise value of $\delta'(s_0)$.

\subsection{Implantation of a resonance: $f\in C_1^+$,
$f'\in C_2^+$}

We corroborate our last statement by examining another Ansatz for $D_1$ in the
context of the last subsection. We replace (\ref{3fourt}) by
\begin{equation}\label{3twthree}
D_1(s)={1\over \alpha}(s-s_1)\, ,
\end{equation}
and again require that $1/f'$ obtained from Eqs.~(\ref{3four}) and
(\ref{3twelve}) be regular at $s_p$ and have a pole at $s'_p$ with residue
$r'$, with conditions (\ref{5conditions}). One finds
\bea\label{3twfour}
\alpha &=& \frac{1}{N}\bar{r}\bar{r}'(s'_p-s_p)^2\co\nonumber\\
s_1 &=& \frac{1}{2}(s_p'+s_p) -\frac{1}{2N}(s_p'-s_p)
\left\{\bar{r}'+\bar{r}-\bar{r}'\bar{r}(D_2(s_p')+D_2(s_p))(s_p'-s_p)\right\}
\; ,\nonumber\\
N&=&\bar{r}-\bar{r}'+\bar{r}\bar{r}'(s_p-s_p')
\left\{D_2(s_p')-D_2(s_p)\right\}\per\nonumber\\
\eea
We see that in general, $\alpha = O(\epsilon)$, whereas $s_1$ is
 anywhere on the real
axis.
 Equation~(\ref{3ninet})
is replaced by
\begin{equation}\label{3twfive}
{1\over f'(s)}={1\over f(s)}+{(s-4)H(s)\over (s-s_p)(s-s'_p)}\,{\alpha\over s-
s_1+\alpha D_2(s)}\per
\end{equation}

If $\epsilon$ is small, $1/f'$ has a pole $\bar{s}_1$ near $s_1$ with small
residue, $O(\epsilon)$. This implies that $1/f'$ exhibits a zero near $s_1$
which can preclude the Ansatz~(\ref{3twthree}). The discussion becomes
delicate if $s_1$ is close to 4 or $s_0$ and the next statements are valid if
$\bar{s}_1$ is outside $O(\epsilon)$ neighborhoods of 4 and $s_0$:
\begin{enumerate}
\item[i)] if $\bar{s}_1<4$ or $\bar{s}_1>s_0$, $f'$ has a pole on the real
axis near $\bar{s}_1$;
\item[ii)] if $4<\bar{s}_1<s_0$, $f'$ has a pair of complex conjugate poles
close to $\bar{s}_1$. These poles are in the first sheet of the branch point
$s=4$ if $\alpha<0$; they are in the second sheet if $\alpha >0$.
\end{enumerate}
We see that the Ansatz~(\ref{3twthree}) has to be rejected for pairs
$(s'_p,r')$ such that $\bar{s}_1<4$ or $\bar{s}_1>s_0$. It is also inadequate
if $4<\bar{s}_1<s_0$ and $\alpha <0$.
 These
requirements on $\alpha$ and $\bar{s}_1$ are fulfilled in a
wedge--shaped domain of
the $(s'_p,r')$-plane, with apex at $(s_p,r)$.

\begin{figure}[htbp]
\begin{center}
\epsfig{file=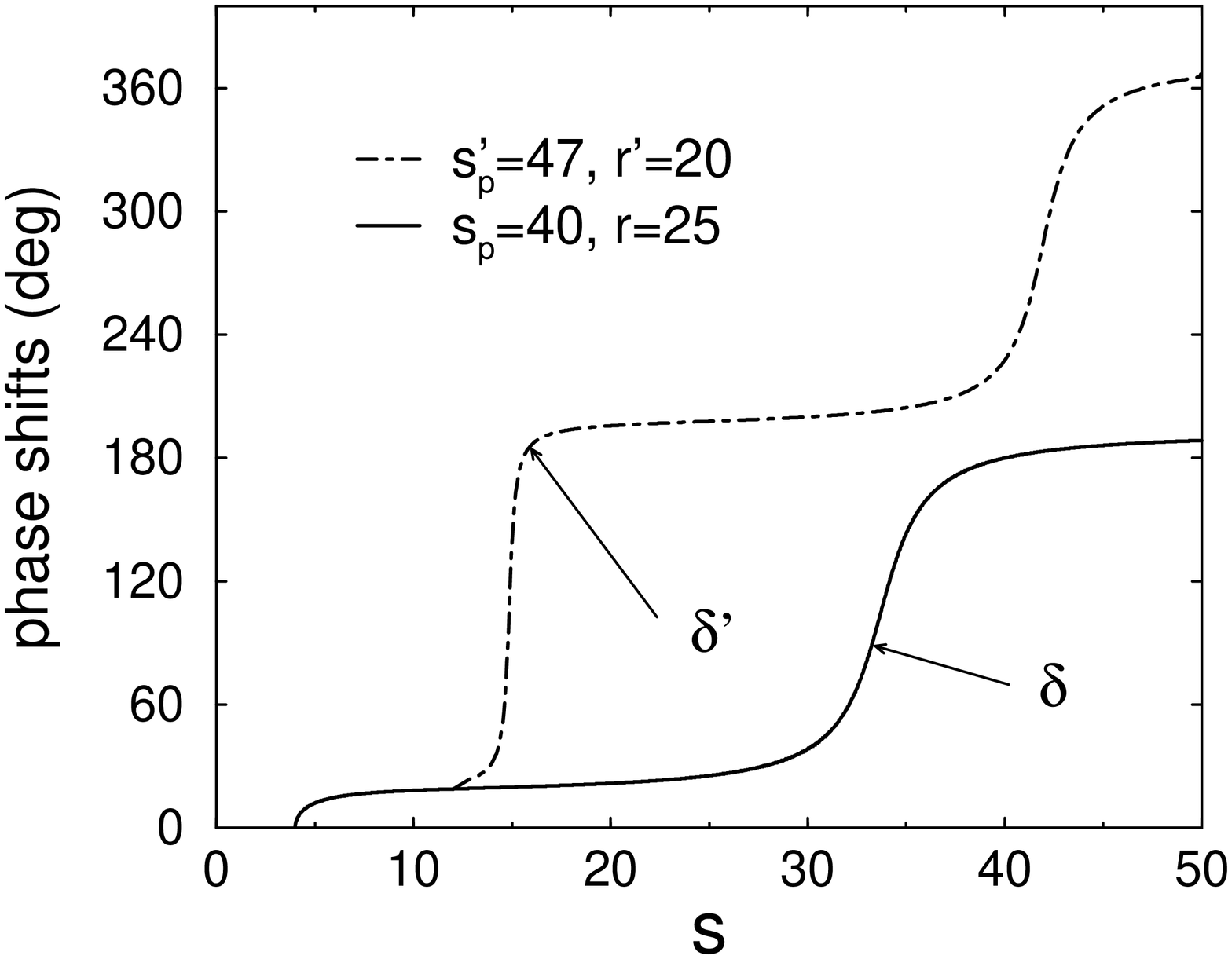,width=8cm,angle=-90}
 \caption{The phase shifts  corresponding to a shift  $C_1^+ \rightarrow
 C_2^+$. The solid
 line is the  phase shift $\delta$ of the function $f$ shown in
Fig.~\protect{\ref{fig3}}, whereas
 the dot--dashed line displays the phase shift $\delta'$ of $f'$, evaluated
 from (\protect{\ref{3twfive}}). The parameters $\alpha$ and $s_1$ are
 evaluated from (\protect{\ref{3twfour}}),  with \protect{(\ref{5new})}.
\label{fig5}}

\epsfig{file=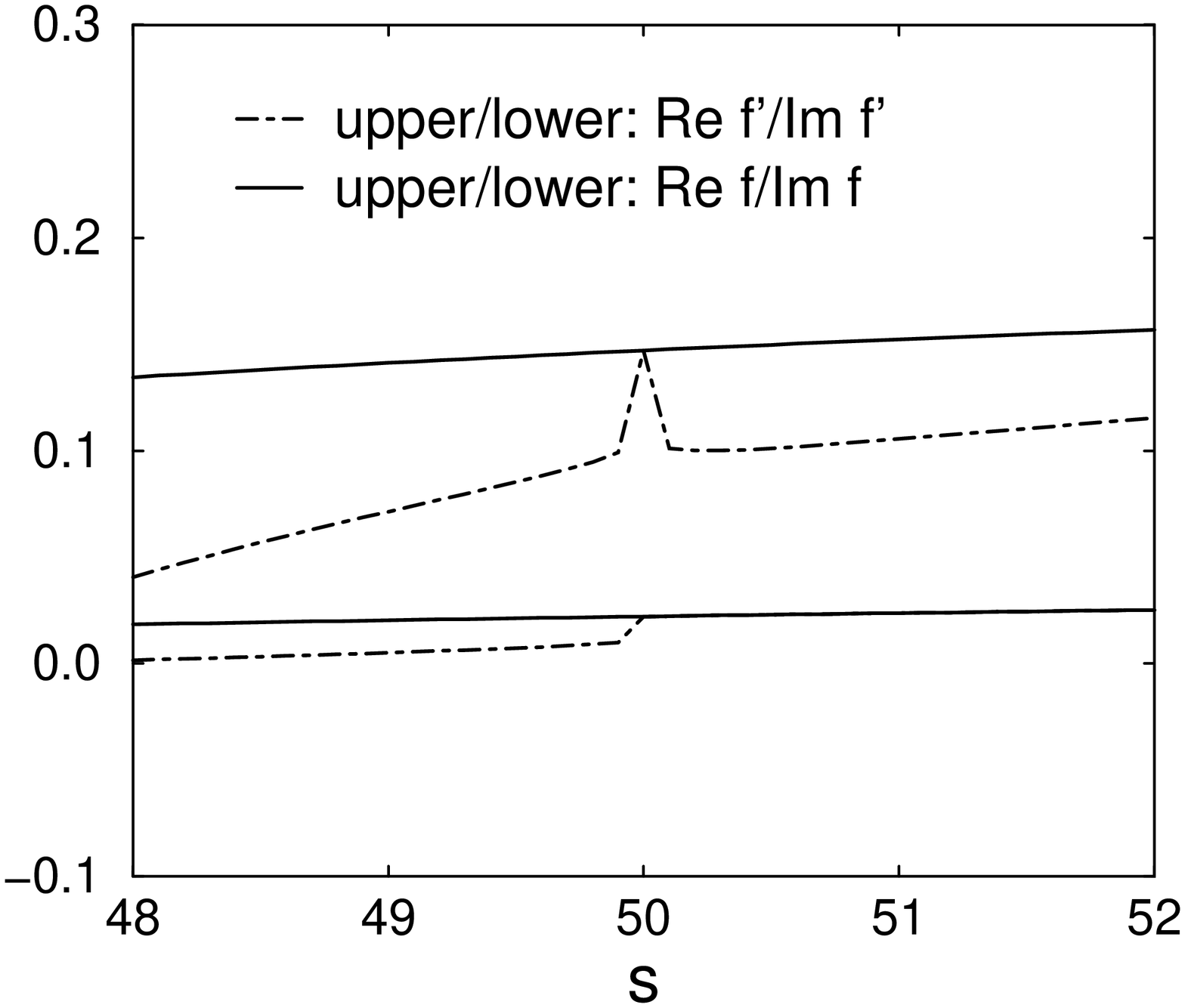,width=8cm,angle=-90}
 \caption{The cusps in $f'$. We display the real and
imaginary parts of $f$ and
$f'$, for the situation
 displayed in Fig.~\protect{\ref{fig5}}.
\label{fig6}}
\end{center}
\end{figure}

If $\alpha>0$, $f'$ generates a solution of
the Roy equation in class $C_2^+$ ($\delta'(s_0)=\delta(s_0)+\pi$). This
solution displays two resonances on $[4,s_0]$: a shifted resonance,
$(r,s_p)\to(r',s'_p)$, and an implanted narrow resonance near $s_1$.
We illustrate this transition in figure~\ref{fig5}.
There, we display the phase shifts  of $f$
 and  $f'$ with a solid and a dot--dashed line, respectively, for
\bea\label{5new}
  s_p'=47\scs r'=20 \scs s_0=50\per
\eea
These parameters are  again outside the linear regime.  The
 difference of the phase shifts  is exactly $\pi$ at the matching point
$s_0=50$. This amounts to  a special case where the CDD pole at $s_p$ is
replaced by two poles at $s_p'$ and  $\bar{s}_1$.

%%%%%%%%%%%%%%%%%%%%%%%%%%

We have pointed out in section 3 that the new solutions exhibit a
singular behaviour at the matching point $s_0$. We illustrate this
feature in figure~\ref{fig6}, where we display the real and the
imaginary part of $ f$ and of $ f'$ with a solid and a
dot--dashed line, respectively, for the situation displayed
 in figure~\ref{fig5}. The singular behaviour of $f'$ is manifest, both in
 the real and
in the imaginary part. Note that, since the imaginary parts of $f$ and
$f'$ agree
 for $s\geq s_0$, the real parts also agree at $s=s_0$ due to
unitarity.
Above the matching point, they are, however, in general different.
Because the real  part of $f$ is
positive in the present case, the figure shows that $f'$ is inside the Argand
diagram, at least in an interval  above $s_0$.

%\end{document}

\subsection{Shift of a resonance: $f,f'\in C_1^-$}
In this example we bring $f$ in (\ref{3ten}) into $C_1^-$ by pushing
$s_0$ below $s_p$ in such a manner that the resonance position is
still below $s_0$, i.e. $\pi/2<\delta(s_0)<\pi$. [The pole at $s_p$ is
 no longer a CDD pole.] The function $H$ used
in the previous two subsections has now to be replaced by a new one
that we call $\hat{H}$. It is given by (\ref{3five}), with $m=1$.
Correspondingly, $\mu$ defined in (\ref{3eight}) becomes $\hat{\mu}$.
It behaves as $s^2$ at infinity and 3 subtractions are needed in the
construction of $D_2$. A convenient redefinition of $D_{1,2}$ in
(\ref{3seven}) leads to
\begin{equation}\label{3twseven}
D(s)=(s-s_p)\left[D_1(s)+(s-s_p)D_2(s)\right]\co
\end{equation}
where
\begin{equation}\label{3tweight}
D_2(s)=s{1\over\pi}\int_{s_0}^\infty{{\rm d}x\over x}{\hat{\mu}(x)\over (x-
s_p)^2}{1\over x-s}\per
\end{equation}
The integral converges because $\hat{\mu}$ has a second--order zero at $s_p$.
As before, we require $1/f'$ to be regular at $s_p$. This implies
\begin{equation}\label{3twnine}
\hat{r}+{\hat{H}(s_p)\over D_1(s_p)}=0\co
\end{equation}
with $\hat{r}=r/(s_p-4)$.
This equation is consistent with real values of $\hat{r}$ and $D_1(s_p)$,
because $\hat{H}(s_p)$ is real. A suitable Ansatz for $D_1$ is
\begin{equation}\label{3th}
D_1(s)=-{\hat{H}(s_p)\over \hat{r}(s_p-s'_p)}(s-s'_p).
\end{equation}
Condition~(\ref{3twnine}) is fulfilled with $s'_p$ a free parameter,
$s'_p>s_0$. Insertion of Eqs.~(\ref{3twnine}) and (\ref{3th}) into
Eq.~(\ref{3four}) gives
\begin{eqnarray}
{1\over f'(s)}&=& {1\over a}+(s-4)\hat{r}\left\{\left[1-{\hat{H}(s)\over
\hat{H}(s_p)}\right]{1\over s-s_p}\right.\nonumber\\[-2mm]
&&  \label{3thone}\\[-2mm]
&& \quad \left.+{\hat{H}(s)\over \hat{H}(s_p)}{1\over s-s'_p+L(s)}\right\}-
\rho(s)\co\nonumber
\end{eqnarray}
where
\begin{equation}\label{3thtwo}
L(s)=\bar{r}{(s'_p-s_p)^2D_2(s)\over 1+\bar{r}(s'_p-s_p)D_2(s)}\per
\end{equation}
Using $\epsilon$ defined in (\ref{3sevent}), we have $L=O(\epsilon^2)$
as long as $s$ is not too large.
For such values of $s$,
$1/f'$ is close to $1/f$ outside a neighborhood of $s_p$ and $s'_p$. We notice
that $\Im\,L_+\neq 0$ for $s>s_0$ and $1/f'$ has no real
pole whose real residue
$r'$ could be imposed (there is a pair of complex conjugate poles near $s'_p$
located in higher sheets of the logarithmic branch point at $s_0$). One finds
that $1/f'$ has no unwanted zeros for any $s$ if $\epsilon$
is small enough, and we end up
with a family of solutions of the Roy equation in $C_1^-$ indexed by the
single parameter $s'_p$. This is in accordance with section~3. The shift of
the resonance position and the change of its width are now correlated and fixed
by the value of $(s'_p-s_p)$. Both $f$ and $f'$ are without CDD poles,
and the non--uniqueness is due to the fact that $\pi/2 < \delta(s_0)
< \pi$.

\setcounter{equation}{0}

%%%%%%%%%%%%%%%%%%%%%%%%%%
%chapter 6
%%%%%%%%%%%%%%%%%%%%%%%%%%%

\section{How to ensure uniqueness}

The solution of the Roy equation
is not unique -- the behavior of a partial wave below $s_0$ cannot be predicted
in a unique way from an input $(a,A)$. Information on the value of the phase
shift at $s_0$ is needed and free parameters have to be fixed if
$\delta(s_0)>\pi/2$.
This non--uniqueness restricts severely the efficiency of a Roy equation
as a tool for the construction of a low--energy extrapolation. We show
in the following,  how uniqueness can be restored in
principle by imposing additional physical requirements. The unphysical
singularity at the matching point is removed at the same time, as it
has to be.

\subsection{Examples of unique solutions}
We start with three illustrative examples.

\vskip.3cm

{\underline{{\it Example 1}}}
In section 5 we have constructed, starting from the amplitude
(\ref{3ten}), several  functions $f'$  that satisfy the Roy
equation with matching point
 $s_0=50$
 and   input generated by $f$,
\bea\label{eq61}
a=f(4)\sem A(s)= \Im f(s)\scs s  \geq s_0\, .
\eea
Two cases are displayed -- in terms of their phase shifts
$\delta'$ -- in Figs.~\ref{fig4} and \ref{fig5}.
 As we already
 mentioned, these phase shifts develop singularities at the matching
point, see Fig.~\ref{fig6}.
 Below  we will show that,
 had we required the new solution $f'$ to be regular at $s_0$, there
would be exactly one solution of the Roy equation with input
(\ref{eq61}),
 namely $f$ itself. In other
words, all solutions $f'\neq f$  develop singularities at $s_0$.
It turns out that this property of the input (\ref{eq61}) is due to
the fact that
the amplitude is elastic above the matching point. Whereas $f$ is
elastic on the whole interval $[4,\infty)$, the following example
shows that a finite interval containing $[4,s_0]$ suffices to render
the solution unique.

\vskip.3cm

{\underline{{\it Example 2}}}
 We  consider the function $f'$ constructed in subsection 5.2, see
Eqs.~(\ref{3twfive}), (\ref{5new}) and Fig.~\ref{fig5}, dot-dashed
line. Suppose we wish to   construct solutions of the Roy equation with
 displaced matching point  $s_0'=45$ and input defined by
the scattering length and
the absorptive part of  $f'$,
\bea\label{eq62}
a=f'(4)\, ; \, A(s)=\Im f'(s) \, , s \geq s_0'\, .
\eea
In the language of section 2, this problem belongs to  class $C_2^-$: the
phase shift at the matching point is $3\pi/2 < \delta'(s_0')<
2\pi$, see Fig.~\ref{fig5}, and according to proposition 1 in section
3, the Roy equation with input (\ref{eq62}) has therefore a 3--dimensional
manifold of solutions.
The uniqueness  statement in example 1 is also true here: Suppose
we seek for solutions of
the Roy equation with input (\ref{eq62}) and require that the solution
is regular at $s=s_0'$. There is again exactly one solution,
namely$f'$ itself.

It is obvious that we are dealing with special inputs --
 we call them  {\it analytic inputs} below. The following example
 displays an input that is not analytic.

\vskip.3cm

{\underline{{\it Example 3}}}
Consider again the
 amplitude (\ref{3ten}). We set the residue $r$ to zero,
\bea
a=f(4)+\epsilon\,  ; \, A(s)=f(s)\, , \, s\geq s_0\, ; \, r=0\, ,
\eea
 with  $s_0=50$. For sufficiently small $\epsilon\neq 0$,
one can show that there is {\it no} solution
that is analytic at $s_0$, see subsection 6.3.

\subsection{Analytic input and uniqueness}
To generalize our findings, we exploit smoothness properties of the
amplitudes which are not explicit in the Roy equations.
 The partial wave
amplitudes enter  this equation as boundary values of analytic
functions. Boundary values need not be smooth, and this is compatible with the
dispersion relation~(\ref{2one}).
 However, it is  quite remarkable
that smoothness  is imposed by elastic unitarity:
\begin{prop}\label{prop3}
Let f be regular in the complex $s$--plane, cut along the real axis
for $s\geq 4$, and let its boundary value $f_+$ verify elastic
unitarity on $[4,s_1]$. Then the real and imaginary parts of
$f_+$ are separately holomorphic  in a
 complex neighborhood of $(4,s_1)$.
\end{prop}

A proof of the proposition is given in \cite{zimmermann}.
Notice that the parameter $s_1$ need not coincide with the first
inelastic threshold $s_{inel}$ -- the proposition is true for any
$s_1 > 4$. Taking $s_1=s_0$, one concludes that
 all solutions $\delta$ of the Roy equation (\ref{2twenty})  are
 regular in the  interval $(4,s_0)$.

Consider  amplitudes that satisfy the conditions of
  proposition 3
 and in addition verify the Roy
 equation with input $(a_f,A_f(s)) = (f(4),\Im f_+(s))$
 for some $s_0 < s_1$.
 The proposition tells us that $f$ is regular at $s_0$.
 A second solution of the same
 equation may be  singular at $s_0$, because it need
 not be elastic on $[s_0,s_1]$. The following proposition shows
  that $f$ is in fact the only
solution which is regular at  $s_0$. This is an important feature,
allowing one in
principle to identify the physical extrapolation within the manifold
of solutions.

\begin{prop}\label{prop4}
Let $f$ be an amplitude that satisfies the conditions of proposition
3  and  that furthermore verifies the Roy equation with
input $(a_f,A_f)$ for some $s_0 <
s_1$. Let $f'\neq f$ be a second solution with the same input. Then $f$ is
regular at $s_0$, whereas $f'$ is singular.
\end{prop}

 We relegate a proof of the proposition to
appendix C. It elaborates observations made in \cite{pomponiuw} and
\cite{epelew}. The examples 1 and 2 given above fulfill the conditions
of the proposition  with $s_1=\infty$  and $s_1=50$,
respectively. Therefore, according to proposition 4,
there is exactly one solution of the Roy equation
 that is regular at the matching
point.

Proposition 4 tells us that there is a special class of inputs that
allow a unique solution that is regular at the matching point $s_0$,
as announced in section 2. According to proposition 3, the
high--energy absorptive part of an input belonging to that class has an
analytic continuation from $[s_0,s_1)$ into a complex neighborhood of
$(4,s_1)$. For this reason, we say that the members of our special
class are {\it analytic inputs}. A physical amplitude $f$ defines an
analytic input $(a_f,A_f)$ if $s_0 < s_{inel}$. Uniqueness is achieved
in the sense that the corresponding Roy equation has exactly one
solution, coinciding with $f$, which is regular at $s_0$.

The existence of a class of inputs ensuring uniqueness and regularity
at the matching point  has been established
in an indirect way. We have no direct
and complete characterization of an analytic input. There are involved
constraints apart from analyticity of the high--energy absorptive part
$A$, and one cannot decide directly if a given input is an analytic
one. In particular, the scattering length $a$ is fixed by $A$ and is
not an independent parameter, see example 3 above. We arrived at a
unique solution of the Roy equation by choosing inputs which are
compatible with elastic unitarity above the matching point. In the
physical context, this means $s_{inel} > s_0$. One can prove that
uniqueness is also obtained if $s_0> s_{inel}$, provided that
 the inelasticity is  sufficiently smooth.

\subsection{Approximate  input and cusps}
Although we are working here with model amplitudes without left--hand
 cut, the results of subsection 6.2 hold true in the physically
 realistic situation with left--hand cut. As a physically relevant
 input is analytic, with $s_1=s_{inel}$, we conclude that it allows a
 unique solution regular at $s_0$ if $s_0 < s_{inel}$. However, a
 physical input is only approximately known, and  one is
 faced in practice
 with  an arbitrary input, as a result of which
  non--uniqueness and unphysical singularities do occur. This is the
reason why we have analyzed in detail the Roy equation in its general
setting.

\begin{figure}[h]
\begin{center}
\epsfig{file=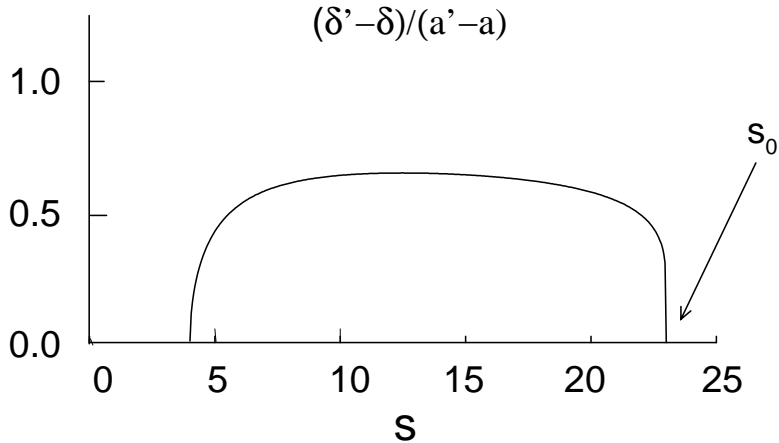,width=6cm,angle=-90}
 \caption{The quantity $(\delta'-\delta)/(a'-a)$ according to
 Eq.~(\protect{\ref{new61}}). The reference amplitude $f$ is the one in
 (\ref{3ten}), and the  parameters used are given  in (\protect{\ref{new62}}).
The cusp generated at $s_0$ is very clearly seen.
\label{fig7}}
\end{center}
\end{figure}

If we know that a high--energy absorptive part $A$ belongs to an
analytic input, but the corresponding scattering length is not
precisely known, one has to work with a trial input $(a',A)$, with
$a'\neq a$.
 We expect that all the
resulting solutions  will have cusps at $s_0$. This can be
established if $a'$ is infinitesimally close to $a$ by using the
techniques developed in~\cite{pomponiuw,epelew}. For instance, if
$\delta(s_0)<\pi/2$ and
$\delta'(s_0)=\delta(s_0)$, the neighboring solution $f'$ is unique and its
phase shift is given by
\begin{equation}\label{new61}
\delta'(s)=\delta(s)+\sigma(s){G(s)\over G(4)}(a'-a)\, ,
\end{equation}
with $G$ defined in (\ref{2fourt}). The ratio
$(\delta'-\delta)/(a'-a)$ is shown in Fig.~\ref{fig7} for a
 neighboring solution of the amplitude $f$ in (\ref{3ten}), using
\bea\label{new62}
s_p=40\scs r=100\scs s_0=23\per
\eea
The cusp at $s_0$ is clearly visible and the
effect of a modified
scattering length extends over the whole interval $(4,s_0)$.
%%%%%%%%%%%%%%%%%%%%%%%%%%
 If an incorrect value $a'$ of
the scattering length also entails a cusp when it is not close to $a$, the
correct scattering length is specified by the fact that it defines an input
allowing a unique solution without a cusp: one solves the Roy equation
with several trial $a'$, and uses
\bea
a\doteq a'_{no cusp}\, .
\eea
 In this sense the scattering length
of a physically realistic  amplitude can be predicted in principle.
However,
even if we are sure that the high--energy absorptive
part is analytic, we do not know
it exactly in practice, and we have to use an approximate form $A'$. We have to
work with an input $(a',A')$ which is meant to approximate the analytic
$(a,A)$. It is unlikely that there will still be a value of $a'$ which removes
the cusp in one of the solutions of the resulting Roy equation. What one can
try in practice is to find the value of $a'$ which minimizes the size of the
cusp.
In the coupled channel case, an analogous procedure may be  used to avoid
solutions that generate a cusp \cite{coletal}.

\setcounter{equation}{0}

%%%%%%%%%%%%%%%%%%%%%%%%%%
%chapter 7
%%%%%%%%%%%%%%%%%%%%%%%%%%%

\section{Summary and conclusions}

The following points summarize the content of this  article.

\begin{enumerate}
\item
In view of forthcoming applications \cite{coletal} of the Roy
equations \cite{roy} in
 the analy\-sis of $K_{l4}$ decays,
 we have considered here
 the one--channel Roy equation.
  We have analyzed the multiplicity and  singularity structure of
  its solutions
 for a given input $(a,A)$ of scattering length $a$ and high--energy
absorptive part $A$.

\item
First, we have  investigated the  infinitesimal neighborhood
of a given solution $\delta$ \cite{pomponiuw}.
 According to proposition \ref{prop1} in section 3, this
neighborhood  contains an
$m$--parameter family of solutions, where
\begin{equation}
m=\left\{\begin{array}{cl}
\left[2\delta(s_0)\over \pi\right] \quad&\mbox{if }
\delta(s_0)>{\pi\over
2},\\[2mm]
0 & \mbox{if }-{\pi\over 2}<\delta(s_0) < {\pi\over 2}.
\end{array}\right.
\end{equation}
The symbol $[x]$ denotes the greatest integer not exceeding $x$.
For a monotonically increasing phase, $m$ counts the number of times
$\delta(s)$ goes through multiples of $\pi/2$ as $s$ varies from
threshold to the matching point $s_0$.
This result illustrates that a given input $(a,A)$ does not, in
general, fix the solution uniquely. One has in addition to fix the
phase shift at the matching point, and to determine the corresponding
$m$ parameters by other means.
\item
Using proposition 2 of section 4, we have  constructed in
 section 5 -- starting from
a given solution $f$
with input
$(a_f,A_f)$ --  additional exact solutions $f'\neq f$ with the same
input.
The function $f'$ contains in general  several arbitrary parameters
that may be used to
either change the position and residues of the poles present in the
inverse amplitude $1/f$,  or to remove (or implement new)  poles.
This  illustrates  that a given input  allows
for solutions with a different value of the phase shift at the matching point
 -- these phase shifts only have to satisfy the boundary condition
 (\ref{2six}).

\item
The solutions $f'\neq f$ so constructed  have the property that they
contain a cusp in
the real and imaginary part at the matching point. An example
is displayed in figure~\ref{fig6}.
The origin of these cusps is made clear in proposition \ref{prop4} of
section 6.

\item
In case we know that a given absorptive belongs to an analytic input
 (see section 6 for this notion), we expect on the basis of
proposition \ref{prop4}
that the corresponding scattering length can be determined
 as the one that results in a solution of the
Roy equation without a cusp at $s_0$.

\item
 Propositions 1  and 3  in sections 3
 and 6  were established long ago in Refs.~\cite{pomponiuw} and
 \cite{zimmermann},
respectively.
 As far as we are aware, propositions 2 and 4 in sections 4 and 6 are,
 on the other hand, new results.

\end{enumerate}

In conclusion,
if we know that we are dealing with an analytic input, the
corresponding  amplitude is the unique regular solution of the Roy
equation.
Because we are in general forced to use  an approximate input,
non--uniqueness and
unphysical singularities  do show up.
A visible cusp in a numerical solution of the Roy equation with
$0 <\delta(s_0) < \pi/2$ is a signal that the input is not physical or a
poor approximation of a physical one. The deficiency
 may be hidden
in the scattering length $a$, in the absorptive part $A$, or in both.

\subsection*{Acknowledgements}

We  thank G. Colangelo for collaboration in an early
stage of this work. We have enjoyed useful
discussions  with  H. Leutwyler, and  we are grateful to  D. Atkinson for
sending us  the theses work of T.P. Pool and of A.C. Heemskerk. One of
us (J.G.) thanks the mathematicians J. R\"atz, M. Reimann and
Th. Rychener for useful discussions concerning  several aspects of this work.

\clearpage

\renewcommand{\thesection}{\Alph{section}}
\renewcommand{\theequation}{\Alph{section}.\arabic{equation}}
\renewcommand{\theprop}{\Alph{section}}
\setcounter{section}{0}
\setcounter{equation}{0}

%%%%%%%%%%%%%%%%%%%%%%%%%%
%appa
%%%%%%%%%%%%%%%%%%%%%%%%%%%

\section{Solving the integral equation (\ref{2ten})}

We first specify the regularity properties of the
partial waves and phase
 shifts considered in the text.
 Then we construct the general
solution of the integral equation (\ref{2ten}).

\subsection{H\"older continuity}
The class of H\"older--continuous
functions is the
appropriate space to consider,
because Cauchy--integrals  map this space  essentially into
itself \cite{muskhel}. We explain this notion.

Consider  a complex--valued function $f$ of a real variable
 $x$. The function $f$ is called H\"older--continuous
 in the interval
 $[a,b]$ with exponent $\mu$, where $0<\mu\leq 1$,
if there is a
 constant $C$ such that
\begin{equation}\label{a1}
|f(x)-f(y)|\leq C\,|x-y|^\mu\ \;\; ; \; x,y\in[a,b]\per
\end{equation}
We call these functions $H$--continuous for short, and denote the
 corresponding space with $H^\mu_{a,b}$.

\begin{prop}
 Let $f\in H^\mu_{a,b}$, with $f(a)=f(b)=0$
and $\mu<1$. Then the  function
\bea
g(y)=\Pint_{\hspace{-1.3mm}a}^b dx \frac{f(x)}{x-y}
\eea
is also an element of $H^\mu_{a,b}$, with the same
exponent $\mu$. The same is true for
\bea\label{eqa9}
g_\pm(y)=\lim_{\epsilon\searrow 0}\int_a^b dx
\frac{f(x)}{x-y\mp i\epsilon}\per
\eea
\end{prop}
The proof may be found in \cite{muskhel}.

\subsection{Regularity requirements}

The input absorptive part $A$ in (\ref{2one}) is assumed to be
 bounded and H\"older continuous in any finite interval
$[s_0,s_0']$ above the matching point $s_0$.
 Furthermore, we seek solutions of the Roy equation that are
$H$--continuous in $[4,s_0]$ and have normal threshold behaviour
(which implies $\mu\leq 1/2$):
\bea\label{a2one}
\mbox{i)}&& A\in H^\mu_{s_0,s_0'}\scs s_0'>s_0\sem A \;\; \mbox{is bounded}
\nonumber \\
\mbox{ii)}&& f\in H^\mu_{4,s_0}\nonumber \\
\mbox{iii)}&&f=a+ia^2\sigma (s)+O(s-4)\scs s\searrow 4\per
\eea

The phase shifts  can then  be chosen $H$-continuous as well,
\bea\label{a2two}
\mbox{iv)}&&\delta\in
H^\mu_{4,s_0}\nonumber\\
\mbox{v)}&&\delta=\sigma(s)[a+O(s-4)]\scs s\searrow 4\per
\eea

\subsection{Solution of the integral equation (\ref{2ten})}
We first show that solving the integral equation (\ref{2ten}) with the
boundary condition (\ref{2twelve})
is equivalent \cite{muskhel}  to solving a boundary value
problem known as {\it
Hilbert problem}. We then construct the general solution of the latter.

We  start by introducing  an auxiliary  function
$\Phi$~\cite{pomponiuw,epelew},
\begin{equation}\label{eqh2}
\Phi(z)=(z-4){1\over \pi}\int_4^{s_0}dx{\sin(2\delta(x))h(x)\over x-4}{1\over
x-z}\per
\end{equation}
For convenience, we use in this subsection  the variable $z$ to
indicate complex values of the
variable $s$.
The function $\Phi$ has the following properties if $h$ is a solution of
 (\ref{2ten}) and (\ref{2twelve}):
\begin{enumerate}
\item[i)] $\Phi(z)=\overline{\Phi(\overline{z})}$ is  regular
in $\C\backslash[4,s_0]\per$
\item[ii)] The boundary values $\displaystyle \Phi_\pm(s)=
\lim_{\epsilon\searrow 0}\Phi(s\pm{\rm i}\epsilon)$ are $H$-continuous
 in $[4,s_0]\per$
\item[iii)] $\Phi_+(s)=e^{4i\delta(s)}
\Phi_-(s)\sem s\in [4,s_0]\per$
\item[iv)] $\Phi(4)=0\per$
\item[v)] $\Phi(s_0)=0\per$
\item[vi)] $\Phi$ is bounded at infinity$\per$
\end{enumerate}
Out of these, we only  prove  property ii) -- the
remaining ones are  easy to
verify.
The unknown $h$ in (\ref{eqh2}) is an element of $H^\mu_{4,s_0}$, with
$0<\mu\leq 1/2$.
 This follows from   Eq.~(\ref{a2two}) -- the same equation shows
that $h=O(s-4)$ at $s=4$.
Together with  $h(s_0)=0$ and with  proposition A in subsection A.1,
it follows that
$\Phi_\pm(s) \in H^\mu_{4,s_0}$, with the same exponent $\mu$.

The problem to determine  functions $\Phi(z)$ with i)--vi)  is called
 a {\it (homogeneous) Hilbert problem}.
We conclude that each solution of the original integral equation generates
 via (\ref{eqh2}) a solution of i)--vi).
Vice versa, each solution of the Hilbert problem has a
 representation of the form (\ref{eqh2}), where $h$ is given by
\bea\label{eqh3}
h(s)= e^{-2i\delta(s)}\Phi_+(s)\per
\eea
 Furthermore, the function $h$ is real and  solves the integral
 equation (\ref{2ten}). To prove this,
 we write a  Cauchy representation for
 $[\Phi(z)-\Phi(4)]/(z-4)$, with a path that wraps around the cut $[4,s_0]$.
 We then deform the outer part of the path towards infinity, in such
 a manner that only the integral above and below the cut $[4,s_0]$ survives.
 By use of condition iv), we find
\bea
\Phi(z)=\frac{(z-4)}{2\pi i}\int_4^{s_0}\frac{dx}{x-4}
\frac{\Phi_+(x)-\Phi_-(x)}{x-z}\per
\eea
Using properties i) and iii), the claim  is easily proven, and we conclude
 that
solving (\ref{2ten}) with (\ref{2twelve}) indeed is   equivalent to
 solving the Hilbert problem i)--vi).

It remains to construct
 the general solution of
i)--vi). First, we observe that the Omn\`es--type function \cite{pomponiuw}
\begin{equation}\label{a3b}
\bar{G}(z)={1\over (s_0-z)^m}\exp\left[{2\over
\pi}\int_4^{s_0}dx{\delta(x)\over x-z}\right]\co
\end{equation}
with $m$ defined in (\ref{2fift}), satisfies property~i). The behaviour at
$s_0$ is given by
\begin{equation}\label{a5bis}
\bar{G}(z)\sim(s_0-z)^\gamma, \qquad\gamma={2\over\pi}\delta(s_0)-m\co
\end{equation}
$-1< \gamma <  1$. The $H$-continuity and threshold behavior of $\delta$
imply that $\bar{G}$ satisfies ii) and iii) except, possibly, at $s_0$.
Outside $s_0$, $(s_0-z)^m\bar{G}$ and its boundary values are nonzero and the
function
\begin{equation}\label{a4}
F(z)=(s_0-z){\Phi(z)\over \bar{G}(z)}
\end{equation}
is regular in $\C\backslash\{s_0\}$. For $z\neq s_0$ it is given by its
Laurent series. The principal part of this series is identically zero because
condition v) and Eq.~(\ref{a5bis}) imply that $F(s_0)=0$. $F(z)$ is therefore
an entire function. Condition
 vi) tells us that $F(z)\sim z^{m+1}$ at infinity, as a result of which
\begin{equation}\label{a7}
F(z)=(s_0-z)Q(z),
\end{equation}
where $Q(z)$ is a polynomial of degree $m$.
Condition~iv) imposes $Q(4)=0$,
and this gives $Q=0$ if $m=0$: the conditions i)--vi) allow only for  the
trivial solution in this case. If $m>0$, $Q(z)=(z-4)P(z)$, with $P$ a
polynomial of degree $m-1$. We conclude  that the general solution of
i)--vi) is given by
\bea\label{eqb2}
\Phi(z)=
(z-4)\bar{G}(z)\displaystyle{\sum_{n=0}^{m-1}c_nz^n} \scs  c_n\in \R\scs
\eea
 with $\Phi=0$ for $m=0$. Using (\ref{eqh3}), the
result (\ref{2thirt}) follows.

%%%%%%%%%%%%%%%%%%%%%%%%%%
%appb
%%%%%%%%%%%%%%%%%%%%%%%%%%%

\setcounter{equation}{0}
\section{Connection with the $N/D$ approach}\label{appn/d}
\newcommand{\product}{\prod}
The $N/D$ method \cite{n/d} transforms the nonlinear Roy equation into
a linear $N$ equation, whereas the method displayed in section 4
linearizes the construction of the solution $f'$, once a first
solution $f$ is known. In this appendix, we establish the relationship
between the two methods.

We write the $N/D$ representation  of $f$ as follows,
\bea\label{eqn1}
f(s)=\frac{n(s)}{d(s)}\per
\eea
The $N$--function $n$ is holomorphic  in $\C\backslash[s_0,\infty)$, and
the $D$--function $d$ is holomorphic in $\C\backslash[4,s_0]$, with
possible CDD poles on the cut.
 Similarly, $f'=n'/d'$. We now determine
the relation between the pairs $(n,d)$ and $(n',d')$.

Equation (\ref{3four}) gives
\bea\label{eqn2}
f'(s)=\frac{D(s)}{F(s)}f(s)\co
\eea
where
\bea\label{eqn3}
F(s)=D(s) +(s-4)H(s)f(s)\per
\eea
Whereas $D$ is meromorphic in $\C\backslash[s_0,\infty)$, $F$ is
meromorphic in $\C\backslash[4,s_0]$. This crucial point is a
consequence of Eqs.~(\ref{3five})--(\ref{3eight}), which imply that
$\Im{F_+(s)}$ = $0$ for $s>s_0$. In order to turn (\ref{eqn2}) into a $N/D$
representation, we need the set $\{p_i\}$ of poles of $D$, as well as
the set $\{z_j\}$ of its zeros on $[4,s_0]$, and define
\bea\label{eqn4}
\bar{D}(s)=\product_i(s-p_i)\product_j\frac{1}{s-z_j}D(s)\per
\eea
To simplify the argument, we assume in the following that the
$N$--functions $n$ and $n'$ have no zeros on $[4,s_0]$ --
the discussion
is more involved and the definitions (\ref{eqn5})  have to be modified
if this is not the case.
The $N/D$ representation of $f'$ is then obtained by eliminating $D$
in favour of $\bar{D}$
 in the expression (\ref{eqn2}), and writing
\bea\label{eqn5}
n'(s)=Cn(s)\bar{D}(s)\sem d'(s)=Cd(s)\bar{F}(s)\co
\eea
where $C$ is a constant and where
\bea\label{eqn6}
\bar{F}(s)=\bar{D}(s)+(s-4)H(s)f(s)
\product_j\frac{1}{s-z_j}\product_i(s-p_i)\per
\eea
The functions $n'$ and $d'$ defined in (\ref{eqn5}) have the correct
analyticity properties. Denominator functions are normalized to one at
infinity \cite{n/d}. We have checked in our examples that $\bar{F}$
has a finite, nonzero  limit at infinity. Therefore, $d'$ is
properly normalized
if $C=1/\bar{F}(\infty)$.
For $\delta(s_0)>0$, $d$ can be written as
\bea\label{eqn7}
d(s)=\product_{k=1}^{r}\left(\frac{s-s_0}{s-s_k}\right)
\exp\left[-\frac{1}{\pi}
\int_4^{s_0}dx \frac{\delta(x)}{x-s}\right]\co
\eea
where the $s_k$ are the $r$ CDD poles of $f$,
$s_k\in (4,s_0), r=[\delta(s_0)/\pi].$
We see that $d'$ is finite at $s_k$ if $s_k$ coincides with a zero
$z_j$ of $D$,  and if $\bar{F}(z_j)=0.$
Otherwise, every $s_k$ and $z_j$ is a CDD pole of $f'$.
Within our assumptions on $n$ and $n'$, the fate of the CDD poles is
dictated by the zeros of $D$: these produce new CDD poles in $f'$ or
remove CDD poles present in $f$. The poles of $D$ are points where $f'=f$.

%%%%%%%%%%%%%%%%%%%%%%%%%%
%appc
%%%%%%%%%%%%%%%%%%%%%%%%%%%

\setcounter{equation}{0}
\section {Proof of proposition \ref{prop4}}

\begin{figure}[h]
\begin{center}
\epsfig{file=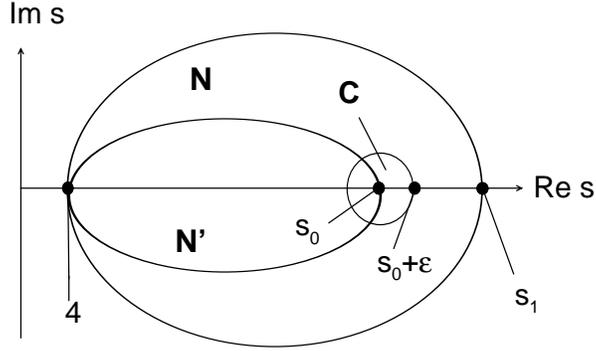,width=4.8cm,angle=-90}
 \caption{Analyticity domains referred to  in the proof of proposition 4.
\label{fig8}}
\end{center}
\end{figure}

Using the framework of section 4,
 we show that
 $f'$ coincides with $f$, if $\Re{f_+'}$ is regular at $s_0$ [in the
sense that it has a  holomorphic extension into a circle of radius
$\epsilon$ and center
$s_0$]. See figure~\ref{fig8} for the  analyticity domains used
in the  proof.

%%%%%%%%%%%%%%%%%%%%%%%%%%

\begin{enumerate}
\item[i)] The dispersion relation (\ref{2one}) written for $f'$ determines
$\Re\,f'_+$ on $\R$. Inversion of this relation gives
\begin{equation}\label{b1}
\Im\,f'_+(s)=-(s-4){1\over \pi}\Pint_{\hspace{-1.3mm}\R}
dx{\phi(x)\over x-s}\co
\end{equation}
where $\phi(s)=(\Re\,f'_+(s)-f'_+(4))/(s-4)$.
The holomorphy of $\Re\,f'_+$ in $C$
implies the holomorphy of $\Im\,f'_+$ in the same circle.
\item[ii)] According to proposition \ref{prop3}, $\Im\,f'_+$ has a
 holomorphic extension into a neighborhood
 $N'$ of $(4,s_0)$, with
$s_0$ on its boundary. Combining this with the previous result we see that
$\Im\,f'_+$ is holomorphic in $\bar{N}=N'\cup C$, a domain extending up to
$s_0+\epsilon$.
\item[iii)]
Proposition 3 tells us that the high--energy absorptive part $A$,
 originally defined on $[s_0,\infty)$, has an analytic
 continuation into a neighborhood N of $(4,s_1)$, which coincides
 with $\Im{f_+}$ on $[4,s_0]$. As $f'$ is a solution of the Roy
 equation with input $(a,A)$, we have
\begin{equation}\label{b2}
\Im\,f'_+(s)=A(s)
\end{equation}
for $s_0\leq s\leq s_0+\epsilon$.
 In view of this equality
and of the regularity of $\Im\,f'_+$ in $\bar{N}$, $\Im\,f'_+$
has to be equal to $A$ on $[4,s_0]$. As $A$ is equal to $\Im\,f_+$ on that
interval, we conclude that $f'=f$.
\item[iv)]
Similarly,
 if $\Im{f_+'}$ is
assumed to be regular at $s_0$, one concludes that
 $\Re{f_+}$ is regular at $s_0$ and $f'=f$.

Therefore, $f'$ has to be singular at $s_0$ if $f'\neq f$. This is the
content of proposition \ref{prop4}.
\end{enumerate}

%%%%%%%%%%%%%%%%%%%%%%%%%%
%bib.tex
%%%%%%%%%%%%%%%%%%%%%%%%%%%

\end{document}